\documentclass[aps,preprint,noshowpacs,nofootinbib,letterpaper,eqsecnum,preprintnumbers, superscriptaddress]{revtex4-1}
\pdfoutput=1
\usepackage{amssymb,amsmath,graphicx}
\usepackage{amsfonts,feynmp}
\usepackage{tipa}
\usepackage{stmaryrd}
\usepackage[pdftex, bookmarks=true,colorlinks,linkcolor=red,urlcolor=blue,citecolor=blue]{hyperref}
\usepackage{color}
\usepackage[bottom]{footmisc}
\newcommand{\non}{\nonumber \\}

      \renewcommand{\th}{\theta}
 
\newcommand{\kap}{\kappa}     
   \newcommand{\sig}{\sigma}
 
   \newcommand{\ome}{\omega}

\newcommand{\Ome}{\Omega}

    \newcommand{\cF}{{\cal F}}

\newcommand{\cO}{{\cal O}}

\newcommand{\rar}{\rightarrow}

\newcommand{\gsim}{ \lower .75ex \hbox{$\sim$} \llap{\raise .27ex \hbox{$>$}} }
\newcommand{\lsim}{ \lower .75ex \hbox{$\sim$} \llap{\raise .27ex \hbox{$<$}} }

\newcommand{\be}{\begin{equation}}
\newcommand{\ee}{\end{equation}}
\newcommand{\bea}{\begin{eqnarray}}
\newcommand{\eea}{\end{eqnarray}}

\begin{document}
\vspace*{.3in} 
\preprint{}
\preprint{IFT-UAM/CSIC-14030}
\vspace*{.3in} 
\title{
BCS instabilities of electron stars to holographic superconductors \\\bigskip
}

\author{Yan Liu} 
\email{{\tt yan.liu@csic.es}}
\affiliation{Instituto de Fisica Teorica UAM/CSIC, C/ Nicolas Cabrera 13-15, Universidad Autonoma de Madrid, Cantoblanco, 28049 Madrid, Spain}
\affiliation{
\mbox{Institute Lorentz for Theoretical Physics, Leiden
  University}\\
\mbox{P.O. Box 9506, Leiden 2300RA, The Netherlands}
}
\author{Koenraad Schalm}
\email{{\tt kschalm@lorentz.leidenuniv.nl}}
\affiliation{
\mbox{Institute Lorentz for Theoretical Physics, Leiden
  University}\\
\mbox{P.O. Box 9506, Leiden 2300RA, The Netherlands}
}
\affiliation{Department of Physics, Harvard University, Cambridge MA 02138, USA\bigskip\bigskip}
\author{Ya-Wen Sun}
\email{{\tt yawen.sun@csic.es}}
\affiliation{Instituto de Fisica Teorica UAM/CSIC, C/ Nicolas Cabrera 13-15, Universidad Autonoma de Madrid, Cantoblanco, 28049 Madrid, Spain}
\affiliation{
\mbox{Institute Lorentz for Theoretical Physics, Leiden
  University}\\
\mbox{P.O. Box 9506, Leiden 2300RA, The Netherlands}
}
\author{Jan Zaanen}
\email{{\tt jan@lorentz.leidenuniv.nl}}
\affiliation{
\mbox{Institute Lorentz for Theoretical Physics, Leiden
  University}\\
\mbox{P.O. Box 9506, Leiden 2300RA, The Netherlands}
}
%\author{Yan Liu, Koenraad Schalm, Ya-Wen Sun and Jan Zaanen}

%\affiliation{Instituto de Fisica Teorica UAM/CSIC, C/ Nicolas Cabrera 13-15, Universidad Autonoma de Madrid, Cantoblanco, 28049 Madrid, Spain}
%\affiliation{Department of Physics, Harvard University, Cambridge MA 02138, USA}
%\affiliation{
%\mbox{Institute Lorentz for Theoretical Physics, Leiden
%  University}\\
%\mbox{P.O. Box 9506, Leiden 2300RA, The Netherlands}
%}

%\email{ {\tt yan.liu, yawen.sun@csic.es, kschalm, jan@lorentz.leidenuniv.nl.}}

\begin{abstract}
\bigskip
We study fermion pairing and condensation towards an ordered state in strongly coupled
quantum critical systems with a holographic AdS/CFT dual. On the
gravity side this is modeled by a system of charged fermion interacting through a BCS coupling. At finite density
such a system has a BCS instability. We combine the relativistic
version of mean-field BCS with the semi-classical fluid approximation
for the many-body state of fermions. The resulting groundstate is 
the
AdS equivalent of a charged neutron star with a superconducting
core% \commentr{\comment{DROP:} and it is shown to be a more stable ground state than the electron star}
. The spectral function of the fermions confirms that the ground
state is ordered through the condensation of the pair operator. 
A natural variant of the BCS star is shown to exist where the gap field
couples Stueckelberg-like to the AdS Maxwell field. This enhances the
tendency of the system to superconduct.

\vspace*{.3in} 

\end{abstract}

\pacs{??.??}

\maketitle

\newpage
%%%%%%%%%%%
%\tableofcontents

%%%%%%%%%%%%%%%%%%%%%%%%%%%%%%%%
%%%%%%%%%%%%%%%%%%%%%%%%%%%%%%%%
\section{Introduction}
%%%%%%%%%%%%%%%%%%%%%%%%%%%%%%%%
%%%%%%%%%%%%%%%%%%%%%%%%%%%%%%%%

Gauge/gravity duality has given us a number of qualitatively new insights into the physics of quantum critical systems. Notably these include a controlled theoretical framework for non-Fermi liquids \cite{Liu:2009dm,Cubrovic:2009ye,Faulkner:2009wj,Faulkner:2010tq} as well as an onset towards superconductivity that is distinct from BCS and  goes beyond Landau-Ginzburg \cite{Hartnoll:2008vx,Hartnoll:2008kx,She:2011cm}. (See e.g. \cite{Iqbal:2011ae} for a % nice
review.) The obvious candidates where both phenomena are seen experimentally are the unconventional high $T_c$ superconductors, and one has reason to hope that gauge-string duality may be able to explain some its open mysteries. 

The clearest puzzle that must be solved to do so, is that one needs a single
holographic model that describes both the non-Fermi-liquid metals and high $T_c$ superconductors
simultaneously. Intuitively this sounds obvious, as the sole charge
carriers are the fermionic electrons; it is their behavior which becomes non-Fermi
liquid-like, while they are simultaneously responsible for the onset of
superconductivity through $d$-wave
pairing. This intuition should not be taken as holy, however. At strong
coupling by definition the underlying electron picture fails, and one
should consider a different weakly coupled set of elementary
excitations. In essence this is what gauge-gravity duality can do very
well. For example, in the specific top-down holographic example of $N=2$ SYM
  with flavor, where one knows the
  explicit Lagrangian of the dual CFT, one can construct a holographic
  superconductor where the order parameter is identified with a
  strongly coupled Cooper
  pair of fermionic ``mesino'' fields \cite{erdmenger}.
% Nevertheless the underlying common origin is reflected in the
% fact that a single model should do both, and here gauge-gravity
% duality can still be improved.

In a bottom-up phenomenological direction, early studies
  that combine pairing with ordering are \cite{Faulkner:2009am}
and \cite{Hartman:2010fk} which studied the formation of a gap in
fermion spectral functions in a holographic superconductor groundstate
and the tendency for holographic non-Fermi liquids to pair and
condense. 
In this paper %letter 
we make a simple further step in the
direction. The aim is to phenomenologically describe a holographic model where
fermion pairing is fully responsible for the superconducting
groundstate. We start from a bulk system with only fermionic matter
fields coupled to gravity and Maxwell field. We include an attractive
four point interaction for the bulk fermions and, approximating the
many-body fermions in the fluid limit, we solve this self-gravitating
charged interacting fermi fluid
in an asymptotically AdS background % in this star limit
at zero temperature. Thus in fact the bulk is a fluid of local BCS vacuum states. More complicated versions of this system are known in the astrophysics
community that studies neutron stars with superconducting
cores. % \commentr{\comment{DROP:} The dual operators of these bulk
  % fermions are composite operators instead of fundamental operators,
  % so this bulk system is not dual to a real BCS superconductor.}
% \commentr{\comment{MOVE TO BELOW} The fluid limit makes it practical to extract
% the macroscopic information of the dual state.}
We show % some
evidence that the holographic dual state to the core-superconducting
electron star is % nevertheless
also 
the pairing induced superconducting state. % \commentr{\comment{DROP} and we expect this bulk system will shed some light on the unconventional superconductors whose paring mechanism remain unsolved and probably involve composite operators. \comment{is it correct that in high tc, pairing is possible related to composite operators?}}

In this construction, the advantages are that the fluid limit makes it practical to extract
the macroscopic information of the dual state. Moreover all the charge
is carried by fermions so that the origin of the boundary charged
degrees of freedom is manifest. However, this construction has the
well-known drawback % \comment{DROP: as the
    % electron star}
  of the fluid limit that the fermionic fields are
not visible at the boundary. We can nevertheless still
  discern boundary effects using the charge distribution within the
  star as we will show later. % To avoid this problem,
In a companion
article \cite{inprogress2} one of us will discuss the same system % microscopically
% including the dynamics of a scalar field \cite{inprogress2}
% and
treating the fermions quantum mechanically \cite{Sachdev:2011ze,Allais:2012ye,Allais:2013lha}. 
To place our work in the context of the previous approaches
\cite{Faulkner:2009am,Hartman:2010fk}, we also discuss a
more generalized model which includes an independent charged scalar
field with dynamics. In the star limit, parameters and fields in this
system will get rescaled and not all the terms in the Lagrangian can
be kept at the same time. In particular the kinetic term always vanishes except in the neutral case. In
  addition to the limit where one goes back to the bulk BCS system,
  there exists a more subtle limit, which we also discuss. % We discuss a
% scaling limit with a charged scalar but without the kinetic term of this scalar.

%Combining a charged scalar and charged fermionic operators
%in a holographic dual, one is naturally led to consider a Yukawa
%interaction between them. In the presence of a chemical potential this
%Yukawa interaction will act exactly as a BCS coupling {\em but now for
  %the fields dual to the composite operators in the strongly coupled
  %field theory.} In the limit where the dynamics of the scalar field
%is neglible one in fact has a true bulk BCS system coupled to gravity
%and electromagnetism. More complicated versions of this system are known in the astrophysics
%community that studies neutron stars with superconducting cores. In
%this letter  

% \commentr{\comment{DROP}
% In ordinary holographic superconductors \cite{Hartnoll:2008vx,Hartnoll:2008kx, Gubser:2009cg, She:2011cm}, the normal state is described by the AdS RN balck hole, which corresponds to a holographic non-Fermi liquid. However, there are some known problems with this AdS RN as a non-Fermi liquid. The properties of the dual fermionic systems reply heavily on the properties of the probes and there are a large amount of charged degrees of freedom hidden behind the horizon. Here in this paper, we consider the normal state to be a limit that the BCS coupling goes to zero, i.e. the electron star, in which all the charge is carried by fermions.}

Let us conclude by emphasizing that we will be studying the
  zero-temperature quantum phase transition between the holographic
  dual of the (Russian doll multi-band) Fermi liquid (the electron
  star) and the pair-ordered BCS groundstate (a star with a BCS
  core) as a function of the BCS coupling.\footnote{We leave the finite
  temperature investigation as an interesting open question.} 
In Sec. \ref{sec2} we will first construct our BCS star and show that it is more stable than the electron star solution at zero temperature% in Sec. \ref{sec2}
. In Sec. \ref{sec3} we show % some
evidence that the bulk BCS star system will correspond to a
superconducting state at the boundary. Then we introduce a more
generalized model in Sec. \ref{seccom} and discuss one scaling limit
that is different from the BCS star one. We conclude in Sec. \ref{sec5}% we have conclusions and discussions
.

% This paper is organized as follows. Section \ref{sec2},

% Section \ref{sec3}
% Section \ref{seccom}

%%%%%%%%%%%%%%%%%%%%%%%%%%%%%%%%
%%%%%%%%%%%%%%%%%%%%%%%%%%%%%%%%
\section{A BCS star}\label{sec2}
%%%%%%%%%%%%%%%%%%%%%%%%%%%%%%%%
%%%%%%%%%%%%%%%%%%%%%%%%%%%%%%%%

% \commentr{\comment{DROP}As mentioned in the introduction, we consider a bulk system where all the charged degrees of freedom are fermions and these fermions have a BCS interaction. Before we start our BCS star construction we first review some basics about BCS theory in condensed matter.}

% In this section we first
%We shall mostly
%consider the bulk Lagrangian of a BCS system without the kinetic term
%from bosons. To make comparison with \cite{inprogress2} and
%\cite{Liu:2013yaa,Nitti:2013xaa}, we will discuss the system with
%kinetic terms in Sec. \ref{seccom} and show that in certain scaling
%limits the bluk BCS Lagrangian in this section can be recovered.

BCS theory \cite{bcs} was proposed by Bardeen, Cooper and Schrieffer
in 1957 as the % first theory to explai
explanation of low temperature superconductivity
through the pairing of fermions into a bosonic state which
  subsequently condenses at low temperatures. Starting with a Fermi
liquid, BCS theory introduces an attractive interaction between
fermions at the Fermi surface. This interaction induces an
instability to the formation of Cooper pairs of
fermions. Microscopically this effective attractive
interaction results from exchange of phonons and % exists
is constrained
in a region $(-\omega_D,\omega_D)$ near the Fermi surface $E_F$ or equivalently the chemical potential $\mu$. Here $\omega_D$ is the Debye frequency, a characteristic scale of phonon excitations. The simplest effective (non-relativistic) Hamiltonian describing the physics of a thin shell of states of width $2 \omega_D$ centered
around the Fermi surface can be written as
\be \label{bcsh} H=\sum_{{\bf k}\sigma} \epsilon _{{\bf k}} c^{\dagger} _{{\bf k}\sigma}c_{{\bf k}\sigma}-\frac{\lambda}{V}\sum_{{\bf k},{\bf k}',{\bf q}}c^{\dagger} _{{\bf k}+{\bf q} \uparrow }c^{\dagger}_{-{\bf k} \downarrow}c_{-{\bf k}'+{\bf q}\downarrow}c_{{\bf k}'\uparrow},
\ee
where $\lambda$ is a positive constant, ${\bf k},{\bf k}',{\bf q}$
denote the momentum, $\sigma =\left\{\uparrow,\downarrow\right\}$ denotes the spin and $\epsilon _{\bf k}$ is the kinetic energy of free fermions.

%We can perform a Hubbard-Stratanovich
%decoupling to make the action quadratic in spinors. We define a condensation operator 

Here we couple the relativistic version of the BCS system to gravity. % and 
The bulk gravity system % that
we consider is the % following
Einstein-Maxwell-BCS system:
\bea \mathcal{L}&=&\frac{1}{2\kappa^2}\bigg(R+\frac{6}{L^2}\bigg)-\frac{1}{4e^2}F_{\mu\nu}F^{\mu\nu}+\mathcal{L}_{\text{BCS}} ,
\eea
 where $\kappa$ is the gravitational coupling constant, $e$ is the Maxwell coupling constant. % and
 $\mathcal{L}_{\text{BCS}}$ is the relativistic Lagrangian % of the relativistic
 % generalization 
of the BCS system
 % We can start from the BCS Lagrangian with an attractive four point interaction term as follows 
 \cite{thesis, Hartman:2010fk}, which is a direct generalization of (\ref{bcsh}) % to a relativistic version
\be \mathcal{L}_{\text{BCS}}=-i\bar\Psi(\Gamma^\mu  \mathcal{D}_\mu-m_f)\Psi+\frac{\lambda}{2}(\bar{\Psi}_c\Gamma^5\Psi)^\dagger(\bar{\Psi}_c\Gamma^5\Psi) 
\ee 
where 
\be\bar{\Psi}=\Psi^{\dagger}\Gamma^{\underline{t}},
~~~\mathcal{D}_\mu=\partial_\mu+\frac{1}{4}\omega_{ab
  \mu}\Gamma^{ab}-iqA_\mu.\ee
Here $\lambda$ is a positive coupling constant of mass dimension
$[\lambda]=-2$ and $\Psi_c=C\bar{\Psi}^T$ and the covariant derivative
includes the gauge- and spin-connection.
%\footnote{Note that $\lambda=
%2\eta_5^2/m_{\phi^2}$ in the notation of \cite{inprogress2}.}
 We 
perform % can introduce
 a Hubbard-Stratanovich transformation as in the non-relativistic case
\be\label{deltaeom0}
\Delta=\lambda \bar{\Psi}_c\Gamma^5\Psi,
\ee
% and the BCS action becomes 
to obtain
\be \mathcal{L}_{\text{BCS}}=-i\bar\Psi(\Gamma^\mu  \mathcal{D}_\mu-m_f)\Psi
+\frac{1}{2}\Delta^\dagger \bar{\Psi}_c\Gamma^5\Psi-\frac{1}{2}\Delta \bar{\Psi}\Gamma^5\Psi_c-\frac{1}{2\lambda}|\Delta|^2.\ee
The auxiliary field $\Delta$, also known as the BCS ``gap'', can be seen as the order parameter for
the BCS condensation.  The connection of this system with a kinetic term for
a dynamical scalar 
$\Delta$ % scalar case
will be discussed in Sec. \ref{seccom}.
%The equation of motion is :
%\bea
%R_{\mu\nu}-\frac{1}{2}g_{\mu\nu}R-\frac{3}{\ell^2}g_{\mu\nu}&=&\kappa^2\big[T_{\mu\nu}^{~\text{gauge}}+T_{\mu\nu}^{~\text{fluid}}\big];\nonumber\\
%{\nabla_\mu}F^{\mu}_{~~\nu}&=&-e^2J_\nu^{~\text{fluid}},
%\eea
%where
%\bea
%T_{\mu\nu}^{~\text{gauge}}&=&\frac{1}{e^2}\big(F_{\mu\rho}F_{\nu}^{~\rho}-\frac{1}{4}F^2g_{\mu\nu}\big);\nonumber\\
%T_{\mu\nu}^{~\text{fluid}}
%&=&(\rho+p)u_\mu u_\nu+pg_{\mu\nu},\nonumber\\
%J_{\mu}^{~\text{fluid}}&=&\sigma u_{\mu}
%\eea

%\comment{Under what condition, can T and J be of form with ideal fluid for a microscopic action?}

The equations of motion for this system are
\bea
\label{emcase0}
R_{\mu\nu}-\frac{1}{2}g_{\mu\nu}R-\frac{3}{L^2}g_{\mu\nu}&=&\kappa^2\big[T_{\mu\nu}^{~\text{gauge}}+T_{\mu\nu}^{~\text{BCS}}\big];\nonumber\\
{\nabla_\mu}F^{\mu}_{~~\nu}&=&-e^2J_\nu^{~\text{BCS}},\\ \nonumber
i(\Gamma^{\mu}\mathcal{D}_{\mu}-m)\Psi+\Delta \Gamma^5\Psi_c&=&0,
\eea
where
\bea
T_{\mu\nu}^{~\text{gauge}}&=&\frac{1}{e^2}\big(F_{\mu\rho}F_{\nu}^{~\rho}-\frac{1}{4}F^2g_{\mu\nu}\big),\nonumber\\\label{emcase1}
T_{\mu\nu}^{~\text{BCS}}&=&\frac{1}{2}\langle i\bar\Psi\Gamma_{(\mu}\mathcal{D}_{\nu)}\Psi-i\bar\Psi \overleftarrow{\mathcal{D}}_{(\mu}\Gamma_{\nu)}\Psi\rangle
+g_{\mu\nu}\langle\mathcal{L}_{\text{BCS}}\rangle,
\\\nonumber
J_{\mu}^{~\text{BCS}}&=&-q\langle\bar{\Psi}\Gamma_\mu\Psi\rangle.
 \eea
As in \cite{Liu:2013yaa}, we rescale $q A_{\mu} \to A_{\mu}$ to fix $q=1.$

%%%%%%%%%%%%%%%%%%%%%%%%%%%%%%%%%%%%%%%%%
\subsection{BCS fluid in the bulk}
%%%%%%%%%%%%%%%%%%%%%%%%%%%%%%%%%%%%%%%%%

%\comment{KS: Why no Yukawa coupling to a dynamical scalar? This
 % connects to the hairy ES and reduces to below in the limit
 % $m^2_{\phi}\rar \infty$} 
%\comment{Einstein-Maxwell-BCS system: backreaction of Hartman-%Hartnoll model}

%We will turn on the BCS interaction among the free electron gas. The microscopic theory for the fluid is described by 
%\be \mathcal{L}_{\text{BCS}}=-i\bar\Psi(\Gamma^\mu  \mathcal{D}_\mu-m_f)\Psi+\frac{\lambda}{2}(\bar{\Psi}_c\Gamma^5\Psi)^\dagger(\bar{\Psi}_c\Gamma^5\Psi).
%\ee
%After performing a Hubbard-Stratanovich transformation  
%$$
%\Delta=\lambda \bar{\Psi}_c\Gamma^5\Psi,
%$$
%the BCS action becomes 
%\be \mathcal{L}_{\text{BCS}}=-i\bar\Psi(\Gamma^\mu  \mathcal{D}_\mu-m_f)\Psi
%+\frac{1}{2}\Delta^\dagger \bar{\Psi}_c\Gamma^5\Psi-\frac{1}{2}\Delta \bar{\Psi}\Gamma^5\Psi_c-\frac{1}{2\lambda}|\Delta|^2.\ee

% Note that the scalar field $\Delta$ under consideration here is nondynamical. We ingore its kinetic energy and also the possible interaction with the gauge field. The connection of this system with  the dynamical scalar case will be discussed in Sec. \ref{seccom}.

As in \cite{Hartnoll:2009ns,deBoer:2009wk,Arsiwalla:2010bt,Hartnoll:2010gu}, we
solve this system in the classical limit $\kappa\to 0$, where we
approximate the many-body-fermi system by an effective fluid. % and to
                                % have an effect from the fermions we
                                % are again in the fluid limit of
                                % fermions where a lot of fermions can
                                % have a finite effect to the gravity
                                % even in the limit $\kappa\to 0$. To
                                % get the follows we have assumed
This is consistent in the adiabatic limit, where the variation of the
electrostatic potential (or local chemical potential) and the gap are small: $\partial_r \mu_l\ll \mu_l^2$ and $\partial_r \Delta\ll \Delta^2$.
This adiabatic limit allows a construction of the expectation values
in (\ref{emcase1}) as if the system is in flat spacetime. We % calculate
% by taking
compute the 
expectation values % on the fermion excitations
at a fixed
local chemical potential $\mu_l$ and gap $\Delta$ and then promote
these to slowly varying quantities governed by $A_t(r)$ and
$\Delta(r)$ respectively. Here $r$ is the radial direction of AdS,
encoding the effective energy scale of the dual field theory.

\begin{figure}[h!]
\begin{center}
\begin{tabular}{cc}
\includegraphics[width=0.5\textwidth]{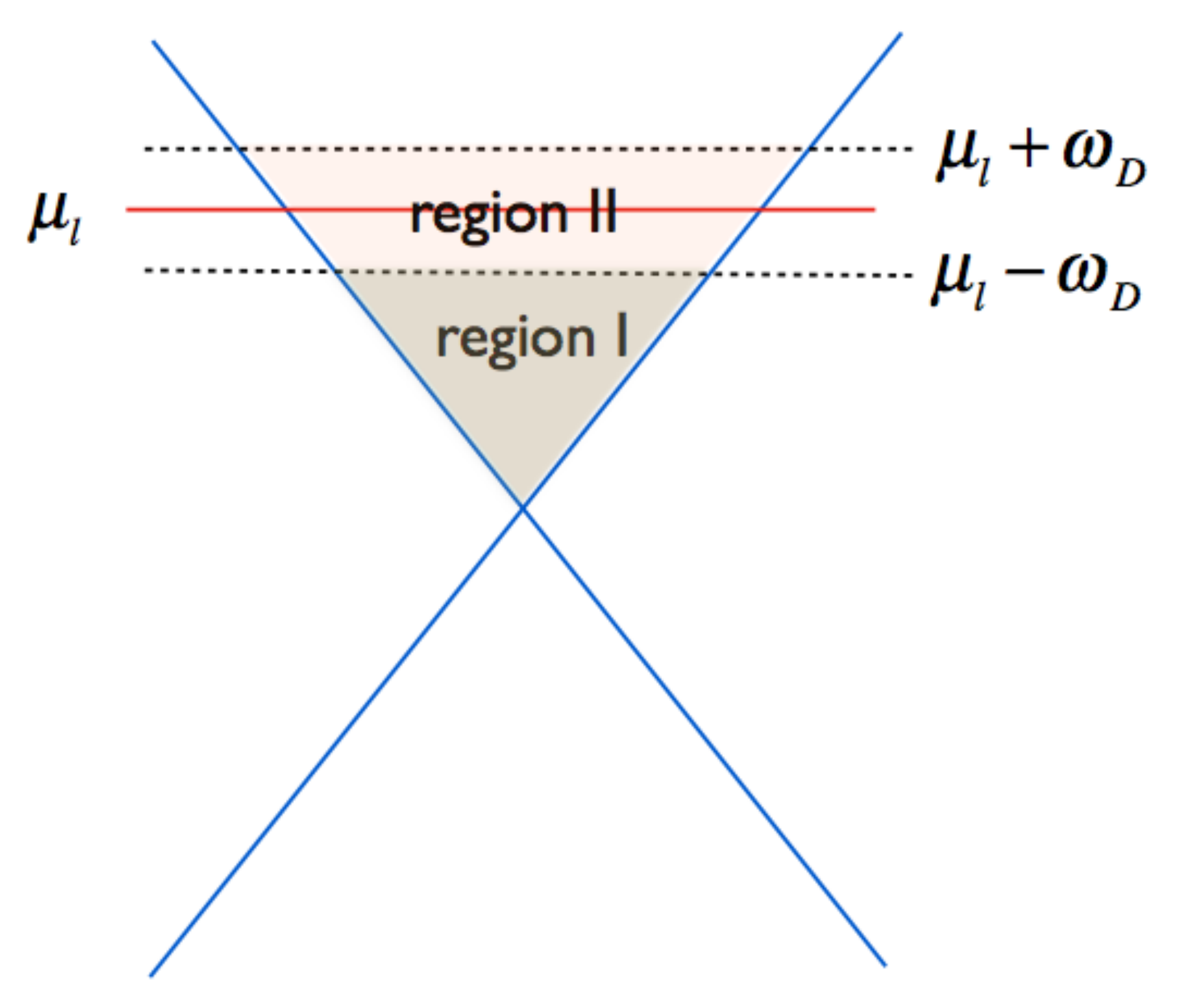}
\end{tabular}
\caption{\small An illustration of the BCS vacuum state. In region I
  the fermions are still free Fermi gas. In region II, the BCS interaction
 allows Cooper pairs start to form and one has a BCS state.}
\label{cartoon}
\end{center}
\end{figure}

To do so remark that
 the BCS interaction term only exists in an interval
 $(-\omega_D,\omega_D)$ near the Fermi surface, so we can divide the
 fermion excitations into two parts, the first part with energy from
 $m_f$ to $\mu_{l}-\omega_D$ and the second part from $\mu_{l}-\omega_D$
 to $\mu_{l}+\omega_D$. This is illustrated in Fig. \ref{cartoon}.
 
%%%%%%%%%%%%%%%%%%%%%%%%%%%%%%%%%%
%%%%%%%%%%%%%%%%%%%%%%%%%%%%%%%%%%

In the first region, the bulk fermion system is still that
  of  free fermions (adiabatically coupled to gravity and electromagnetism) which obey
the Pauli exclusion principle, so it is straightforward to write out the
contribution of fermions in this region to the energy momentum tensor
and the current. They are the regular values for many-body fermions in
the Thomas-Fermi approximation:
\be T_{\mu\nu}^{\text{FL}_{\text{I}}}=(\rho_{\text{I}}^{\text{FL}}+p_{\text{I}}^{\text{FL}})u_{\mu}u_{\nu}+p_{\text{I}}^{\text{FL}} g_{\mu\nu},\ee 
and 
\be J^{\mu}_{\text{FL}_{\text{I}}}= n_{\text{I}}^{\text{FL}} u^{\mu},\ee 
where 
\bea 
\rho_{\text{I}}^{\text{FL}} &=&\int_{{\bf k }^2< \mu_{l}-\omega_D} d^3{\bf k}\langle {\bf k}| T_{00}|{\bf k} \rangle  = \frac{1}{\pi^2}\int_{m_f}^{\mu_l-\omega_D}d\omega \omega^2 \sqrt{\omega^2-m_f^2},  \\ p_{\text{I}}^{\text{FL}}&=&\int _{{\bf k}^2< \mu_{l}-\omega_D} d^3{\bf k}\langle {\bf k}| T_{11}| {\bf k}\rangle = \frac{1}{3\pi^2}\int_{m_f}^{\mu_{l}-\omega_D} d \omega \sqrt{\omega^2-m_f^2}^3,\\  \label{freefermionsecII}
%\sigma_{\text{I}}^{FL}= 
n_{\text{I}}^{\text{FL}}&=&\int_{{\bf k }^2< \mu_{l}-\omega_D}d^3{\bf k}  =\frac{1}{\pi^2} \int_{m_f}^{\mu_{l}-\omega_D} d \omega \omega \sqrt{\omega^2-m_f^2},
\eea 
with $n_{\text{I}}^{\text{FL}}$ denotes the number density of free fermions in region I. 
%Note that before Turning on the BCS interaction, we have free Fermi gas with 
%\be\label{freefg}
%\rho_0=\frac{1}{\pi^2}\int_{m_f}^{\mu_l}d\omega \omega^2 \sqrt{\omega^2-m_f^2},p_0=\frac{1}{3\pi^2}\int_{m_f}^{\mu_{l}} d \omega \sqrt{\omega^2-m_f^2}^3,n_0=\frac{1}{\pi^2} \int_{m_f}^{\mu_{l}} d \omega \omega \sqrt{\omega^2-m_f^2}.
%\ee

In the second region (region II in Fig. \ref{cartoon}), due to the interactions with $\Delta$, fermions do not obey the zero temperature Fermi-Dirac distribution anymore. 
We can first perform a Bogoliubov transformation to make the interacting system tractable. In this interacting region, quasi particles of fermion excitations with opposite momentum and spin near the Fermi surface are coupled together, which introduces off-diagonal elements in the Hamiltonian \cite{thesis} as 
\be\label{hamiltonianbcs} H-\mu_l N= \sum_{\bf k} \Psi^{\dagger}_{\bf
  k} \begin{pmatrix} \xi_{\bf k} & -\Delta \\ -\bar{\Delta} &-\xi_{\bf
    k}  \end{pmatrix}  \Psi_{\bf k} +\sum_{\bf k} \xi _{\bf k}
+V\frac{\Delta^2}{2\lambda} 
\ee
where $\Psi_{\bf k}$ is the Nambu spinor $\Psi_{\bf k}=\begin{pmatrix} c_{{\bf k}\uparrow} \\ 
c^\dagger_{-{\bf k}\downarrow}\end{pmatrix}$, $\xi_{\bf k}$ equals
$\xi _{\bf k}=\epsilon_{\bf k}-\mu_l$, the second term arises from anticommuniting $c^\dagger c$ and $V$ is the volume of the system under consideration.
  
A Bogoliubov transformation can then diagonalize the hamiltonian by redefining
\be  \begin{pmatrix} \alpha_{{\bf k}\uparrow}\\ \alpha_{-{\bf k}\downarrow}^{\dagger}  \end{pmatrix}=\begin{pmatrix} \cos \theta_{\bf k} & \sin \theta_{\bf k}\\ \sin \theta_{\bf k} &-\cos\theta_{\bf k}  \end{pmatrix}  \begin{pmatrix} c_{{\bf k}\uparrow}\\ c_{-{\bf k}\downarrow}^{\dagger}  \end{pmatrix},\ee
 where 
\be \cos (2\theta_{\bf k})=\xi_{\bf k}/E_{\bf k},\ee \be\sin (2\theta_{\bf k})=-\Delta/E_{\bf k},\ee 
and $E_{\bf k}=\sqrt{\Delta^2+\xi_{\bf k}^2}$ is the energy of the
excitations created by $\alpha_{{\bf k}\sigma}^{\dagger}$. 
Note that $\th_{\bf k}$ is such that in the limit $\Delta
  \rar 0$ ($\lambda\rar 0$), it equals $\th_{\bf k}= \pi/2$
for $k<k_F$ and $\th_{\bf k}=0$ for $k>k_F$. After this Bogoliubov transformation, the Hamiltonian becomes 
\be H-\mu_l N=\sum_{{\bf k}\sigma} E_{\bf k} \alpha_{{\bf k}\sigma}^{\dagger}\alpha_{{\bf k}\sigma}+\sum_{{\bf k}}(\xi_{\bf k}-E_{\bf k})+V\frac{\Delta^2}{2\lambda}.
\ee 
The first term in the diagonalized Hamiltonian is related to the
energy of excitations and the rest corresponds to the BCS vacuum
energy, which is the lowest energy state under the BCS interactions. 
% This $\alpha_{{\bf k}\sigma}^{\dagger}$ has a minimum energy of $\Delta$ which
%     means that the Fermi sea annihilated by $\alpha_{{\bf k}}$ has a
%     lower energy than the Fermi sea. Thus 
The BCS groundstate is 
\be \label{bcsvac}
|\Omega_{\text{BCS}}\rangle=\prod_{{\bf k}}\alpha_{{\bf k}\uparrow}\alpha_{-{\bf k}\downarrow}|\Omega\rangle\sim \prod_{{\bf k}}(\cos\theta_{\bf k}-\sin\theta_{\bf k} c_{{\bf k}\uparrow}^{\dagger}c_{-{\bf k}\downarrow}^{\dagger})|\Omega\rangle,
\ee
where $|\Ome\rangle$ is the vacuum state annihilated by $c_{{\bf
  k}\sig}$. Here the range of ${\bf k}$ is within region II in
Fig. \ref{cartoon}. Note that in the
limit $\lambda \rar 0$ (i.e. $\Delta \rar 0$) the BCS vacuum reduces to % the corresponding part of
the Fermi liquid, as $\th_{\bf k}=\pi/2$ for $k<k_F$.

We % can
see that in the limit $\Delta\to 0$, the ground state goes back to the Fermi sea with chemical potential $\mu_l$. For $\Delta$ % becomes
nonzero, Cooper pairs % start to
form and effectively the population number below $\mu_l$
decreases while the population number above $\mu_l$ becomes
nonzero. For small $\Delta$, this BCS vacuum state can be seen as the
resulting state of the free Fermi sea deformed by the BCS
interaction. 
% \commentr{\comment{MOVE TO LATER} For an electron star, there is a filled Fermi sea at each radial slice in the bulk. Here a BCS star has a BCS vacuum at each radial slice in the bulk. The BCS vacuum will go back to the free Fermi sea when $\Delta\to 0$.}

%We set $q=1$.

%$\comment{Two-fluid~model:}$

%Region I: Free Fermion gas + BCS gas\\
%Region II: BCS gas\\
%matching point: $\omega_D=\mu-m$

%${\bf Free~fluid~parameters}$

%\be\label{freefluid} \rho_0=\int_{m_f}^{\mu}d\omega \omega g(\omega),~~
%n_0=\int_{m_f}^{\mu}d\omega g(\omega),~~
%p_0=\mu n_0-\rho_0,\ee
%where $g(\omega)=\frac{1}{\pi^2}\omega\sqrt{\omega^2-m_f^2}.$

 For our purpose, we 
need to compute
 the expectation values of the macroscopic properties $\rho$, $p$ and
 $n$ of the fermion system in the BCS vacuum state in region II. We
can choose the phase of the complex scalar to be zero. The energy
in the BCS vacuum can be directly read from the diagonalized
Hamiltonian. Note that the expression in
 (\ref{hamiltonianbcs}) includes a chemical potential term. We also treat the potential term for $\Delta$ and separately. The energy density of the BCS sector
therefore constitutes of three parts
\begin{align}
  \label{eq:1}
  \rho_{\text{II}} &= \langle \Omega_{\text{BCS}} | T_{00}%\frac{\hat{H}}{V} 
  |\Omega_{\text{BCS}} \rangle \\
&=\langle \Omega_{\text{BCS}} | \frac{1}{V}\bigg(\hat{H} -\mu_l \hat{N} -
V\frac{\Delta^2}{2\lambda}\bigg)|\Omega_{\text{BCS}} \rangle + \mu_l n_{\text{II}} + \rho_{\Delta} 
\end{align}
where $n_{\text{II}}$ is the number (=charge) density
from region $\text{II}$, which we will compute momentarily and
$\rho_{\Delta}=\Delta^2/2\lambda$.
Explicitly the term to be evaluated is
\begin{align}
  \label{eq:2}
  \rho_{\text{II}} % &= \langle \Omega_{BCS} | \hat{H} -\mu \hat{N
 %  } -
% V\frac{\Delta^2}{2\lambda}|\Omega_{BCS} \rangle \non
&= \frac{1}{V}\sum_{\bf k} \left(\xi_{\bf k} - E_{\bf k}\right) + \mu_l n_{\text{II}} + \rho_{\Delta}.
\end{align}
The sum here ranges over all momentum states in region II. To
evaluate it, we note that in the fluid limit, the sum can be
substituted for an integral and change integration variables
\bea
 \rho_{\text{II}} &=& \int_{\text{region}~\text{II}} \frac{d^3k}{(2\pi)^3} \left(\xi_{\bf k} -
E_{\bf k}\right)+ \mu_l n_{\text{II}} + \rho_{\Delta} \nonumber\\
%&=  \int_{\text{region}~\text{II}} \left(\frac{k^2}{2\pi^2}  \frac{dk}{d\xi}\right) d\xi (\xi
%-\sqrt{\xi^2+\Delta^2})+ \mu n_{\text{II}} + \rho_{\Delta} \non
&=&\int_{\text{region}~\text{II}} d\xi \nu(\xi)\bigg(\xi
-\sqrt{\xi^2+\Delta^2}\bigg)+ \mu_l n_{\text{II}} + \rho_{\Delta} 
\eea
where 
\be 
\nu(\xi)=\frac{1}{2\pi^2}(\mu_l+\xi)\sqrt{(\mu_l+\xi)^2-m_f^2}
\ee
is the number density as a function of the effective energy $\xi= \omega({\bf k})-\mu_l$.
It follows directly from the relativistic dispersion relation $\xi =
\sqrt{{\bf k}^2+m_f^2}-\mu_l$. Noting that $\xi$ vanishes at $\mu_l$ the
boundaries of the integral are immediately seen to be,
\footnote{Note that a change of integration variables to the physical energy
$E=\sqrt{\xi^2+\Delta^2}$ exposes the well known gap for
$|E|<|\Delta|$
\be
\rho_{\text{II}} =
%\int_{-\omega_D}^{\omega_D} d\xi \nu(\xi)(\xi
%-\sqrt{\xi^2+\Delta^2}) 
\int_{-\sqrt{\omega_D^2-\Delta^2}}^{\sqrt{\omega_D^2-\Delta^2}} dE \nu_E(E)(\sqrt{E^2-\Delta^2}
-E) + \mu_l n_{\text{II}} + \rho_{\Delta} 
\ee
with 
$$\nu_{E} = \frac{\th(|E|-|\Delta|)}{2\pi^2}\frac{E (\mu_l+\sqrt{E^2-\Delta^2})}{
\sqrt{E^2-\Delta^2}} \sqrt{(\mu_l+\sqrt{E^2-\Delta^2})^2-m_f^2}.
$$
We will use the effective energy $\xi$ instead for convenience.
\par
}
\begin{align}
\label{rho} 
\rho_{\text{II}} &=\int_{-\omega_D}^{\omega_D} d\xi \nu(\xi)\bigg(\xi
-\sqrt{\xi^2+\Delta^2}\bigg) + \mu_l n_{\text{II}} + \rho_{\Delta}.
\end{align}

We similarly compute the total charge density in the BCS state. This
is still measured by the number operator $\hat{n}= \sum_{{\bf{k}}\sig}c_{{\bf
    k}\sig}^{\dagger} c_{{\bf k}\sig}$. One finds 
(the factor 2 is the spin degeneracy)
\bea
  \label{eq:3}
  n_{\text{II}} &=& \langle \Omega_{\text{BCS}}| \hat{n} |\Omega_{\text{BCS}} \rangle = \int_{\text{region}~\text{II}} \frac{d^3k}{(2\pi)^3} 2 (\sin^2 \th_{\bf{k}})  \nonumber\\
&=&\int_{-\omega_D}^{\omega_D} d\xi\nu(\xi) \left(1- \frac{\xi}{\sqrt{\xi^2+\Delta^2}} \right).
\eea
We can see from this expression of the number density that the occupation number for each spin at a momentum below the chemical potential is in a range $1/2$ to $1$ while the occupation number above the chemical potential is smaller than $1/2$. At the chemical potential the occupation number is exactly $1/2$. When $\Delta\to 0$, the occupations numbers will return to that of the free Fermi gas.

The pressure is computed from the expectation value of the spatial
components of the stress tensor in the BCS vacuum
state. 
The expression for the stress tensor is in Eq.~(\ref{emcase1}). Using
isotropy, $\langle T_{ij}\rangle= p\delta_{ij}$, $(i=1,2,3)$, the computation simplifies and we obtain
\begin{align}
\label{p} p_{\text{II}}&=\langle\Omega_{{\text
    {BCS}}}|T_{11}|\Omega_{\text{BCS}}\rangle\non
& =\frac{1}{3}\int_{-\omega_D}^{\omega_D}d\xi
\nu(\xi)\frac{(\mu_l+\xi)^2-m_f^2}{\mu_l+\xi}\bigg(1-\frac{\xi}{\sqrt{\xi^2+|\Delta|^2}}\bigg)
+ p_{\Delta}.
\end{align}
The last term $p_{\Delta}=-\Delta^2/2\lambda$ arises from the
classical term in the Lagrangian (the pure potential contribution).

For calculational convenience we evaluate these expressions in the
limit $\Delta\ll \omega_D\ll \mu_l$ with $m_f\ll\mu_l$ and express
them in terms of the difference of BCS system compared to Fermi liquid
at the same chemical potential. The first inequality is justified as the
self-consistent solution for the gap
$\Delta$ is notoriously exponentially
smaller than the other scales. This is guaranteed if the second
inequality $\ome_D\ll \mu_l$ holds.  We will confirm this momentarily.
The approximation $\Delta \ll \ome_D$ is (well known to be) subtle, because one cannot expand in $\Delta$ in
the integrand. We therefore first use $\ome_D \ll (\mu_l,\mu_l-m_f)$ to approximate the
density of states as
\begin{align}
\nu(\xi)=\frac{1}{2\pi^2}(\mu_l+\xi)\sqrt{(\mu_l+\xi)^2-m_f^2}=\nu_0+\xi \nu_1+% \xi^2\nu_2+
\dots
\end{align}
where
\be\nu_0=\frac{1}{2\pi^2}\mu_l\sqrt{\mu_l^2-m_f^2},~~
\nu_1=\frac{1}{2\pi^2}\frac{2\mu_l^2-m_f^2}{(\mu_l^2-m_f^2)^{1/2}}
%\nu_2=\frac{1}{4\pi^2}\frac{\mu_l(2\mu_l^2-3m_f^2)}{(\mu_l^2-m_f^2)^{3/2}}
.
\ee

We also expand the expression of $ \rho_{\text{II}}$, $n_{\text{II}}$ and $p_{\text{II}}$ in this limit 
\be\label{limit} \Delta\ll \omega_D\ll(\mu_l,\mu_l-m_f)\ee and then
subtract these free fermion contributions from the BCS results. 
This way we isolate the contribution due to the gap
  $\Delta$. % In
% this limit the powers of $\omega_D$ all \commentr{cancel after the
%   subtraction and
We find (see appendix \ref{fluidapp} for details) 
%Reinstating all lower order
%terms in $\nu(\xi)$ they are
%\comment{CHECK this as well as appendix \ref{fluidapp}}
\bea 
\label{IIa}
\rho_{\text{II}}&=&\rho_{\text{II}}^{\text{FL}}+ \rho_\Delta+\frac{1}{2\pi^2}\frac{\mu_l^3}{\sqrt{\mu_l^2-m_f^2}}\Delta^2\ln\frac{2\omega_D}{\Delta}+\dots
\\
\label{IIb}
 n_{\text{II}}&=&n_{\text{II}}^{\text{FL}}+\frac{2\mu_l^2-m_f^2}{2\pi^2\sqrt{\mu_l^2-m_f^2}}\Delta^2\ln\frac{2\omega_D}{\Delta}+\dots
\\
\label{IIc}
 p_{\text{II}}&=&p_{\text{II}}^{\text{FL}}+p_\Delta+\frac{\mu_l\sqrt{\mu_l^2-m_f^2}}{2\pi^2}\Delta^2\ln\frac{2\omega_D}{\Delta}+\dots
\eea
where  
\be 
\label{FLII}
\rho_{\text{II}}^{\text{FL}}=\int_{\mu_l-\omega_D}^{\mu_l}d\omega \omega g(\omega),~~
n_{\text{II}}^{\text{FL}}=\int_{\mu_l-\omega_D}^{\mu_l}d\omega g(\omega),~~
p_{\text{II}}^{\text{FL}}=\frac{1}{3} \int_{\mu_l-\omega_D}^{\mu_l}d\omega
\frac{\omega^2-m_f^2}{\omega} g(\omega)
\ee
are the standard Fermi liquid contributions from region II 
with $g(\omega)=\frac{1}{\pi^2}\omega\sqrt{\omega^2-m_f^2}$;
$\rho_\Delta=-p_\Delta=\frac{\Delta^2}{2\lambda}$ as before, 
and the ``\dots'' are higher order terms in $\Delta/\omega_D$ and $\omega_D/\mu_l$. 
%\commentr
%{
%where $g(\ome)= 2\nu(\ome-\mu)$ includes the spin degeneracy.
%}

Finally we use the equation of motion (\ref{deltaeom0}) for $\Delta$ 
\be\Delta =\lambda \langle\bar{\Psi}_c\Gamma^5\Psi\rangle=
\lambda\int_{-\omega_D}^{\omega_D}d\xi
\nu_l(\xi)\frac{\Delta}{\sqrt{\xi^2+|\Delta|^2}},\ee 
which can be integrated to give  \be
\Delta=\frac{\lambda}{\pi^2}\mu_l\sqrt{\mu_l^2-m_f^2}\Delta\ln\frac{2\omega_D}{\Delta}
\ee 
in the approximation of (\ref{limit}).
This equation can be solved to yield
\be
\label{sec:bcs-fluid-bulk}
\Delta=2\omega_D e^{-1/(2\lambda\nu_0)}.
\ee
This well-known suppression of the gap shows the self-consistency of
the assumption $\Delta\ll \omega_D$ in perturbation theory:
perturbation theory holds when $\lambda \nu_0 \ll 1$%$\lambda \omega_D^2 \ll 1$
; for $\omega_D \ll \mu_l$ this implies %$\lambda \nu_0 \ll 1$. Hence 
$\Delta\ll \omega_D$.

Substituting Eq.~(\ref{sec:bcs-fluid-bulk}) into the expressions for $\rho_{\text{II}},
n_{\text{II}}, p_{\text{II}}$, the terms of order
$\Delta^2$ without a logarithm are subleading for $\lambda\nu_0 \ll
1$. We obtain
\bea
  \label{eq:11}
  n_{\text{II}} &=& n_{\text{II}}^{\text{FL}}+ \delta n,~~~~ \delta n =
  \frac{\nu_1}{\nu_0}\frac{\Delta^2}{2\lambda} ,
%=  \frac{2\mu^2-m^2}{\mu(\mu^2-m_f^2)} \frac{\Delta^2}{2\lambda} 
\nonumber\\
 \rho_{\text{II}} &=& \rho_{\text{II}}^{\text{FL}}+ \delta \rho, ~~~~ \delta \rho =\mu_l \frac{\nu_1}{\nu_0}\frac{\Delta^2}{2\lambda},
%  \frac{2\mu^2-m^2}{(\mu^2-m_f^2)} \frac{\Delta^2}{2\lambda}
\\
 p_{\text{II}} &=& p_{\text{II}}^{\text{FL}}+ \delta p,~~~~ \delta p =
%  \frac{\Delta^2}{2\lambda} -\frac{\Delta^2}{2\lambda}=
  0.\nonumber
\eea

Combining this with the pure Fermi liquid contribution from region I,
the total bulk fluid contribution is 
\be\label{totalfluid}
\rho=\rho^{\text{FL}}+\delta\rho_{\text{total}},~~~p=p^{\text{FL}}+\delta p_{\text{total}}, ~~~n=n^{\text{FL}}+\delta n_{\text{total}}
\ee
where $\rho^{\text{FL}},p^{\text{FL}},n^{\text{FL}}$ are the standard Fermi liquid
densities at finite density $\mu$ and $\delta\rho_{\text{total}},
\delta p_{\text{total}},$ $\delta n_{\text{total}}$ are the expressions in
(\ref{eq:11}). 
Explicitly they are
\bea
\label{deltarho}
\delta\rho_{\text{total}}&=&\frac{2\mu_l^2-m_f^2}{(\mu_l^2-m_f^2)}\frac{\Delta^2}{2\lambda},\\ 
\label{deltan}\delta n_{\text{total}}&=&\frac{2\mu_l^2-m_f^2}{\mu_l(\mu_l^2-m_f^2)}\frac{\Delta^2}{2\lambda},\\
\delta\label{deltap} p_{\text{total}}&=&0.
\eea
 Note that 
the standard equation of state for the whole system is still obeyed  
$$\rho+p=\mu_l n.$$

%%%%%%%%%%%%%%%%%%%%%%%%%%%%%%%
\subsection{BCS star background} 
%%%%%%%%%%%%%%%%%%%%%%%%%%%%%%%

Having obtained the parameters of the effective BCS fluid% in the equations of motion
, we now couple the fluid to AdS-Einstein-Maxwell theory as in
Eq. (\ref{emcase0}) % this system
and search for an asymptotically AdS
solution of a self-gravitating BCS star. % Before that, we
We define the dimensionless variables 
%first rescale these parameters in accord with the $\kappa\to 0$ limit. Rescale
\be\label{rescalesecII}
A=\frac{eL}{\kappa}\hat{A},~~~(\rho, p)=\frac{1}{\kappa^2L^2}(\hat{\rho},\hat{p}),~~~
n=\frac{1}{e\kappa L^2} \hat{n},
%\sigma=\frac{1}{e^2}\hat{\sigma},
~~~\lambda=\frac{e^2L^2}{\beta}\hat{\lambda},\ee
\be(m_f,\mu_{l})=\frac{e}{\kappa}(\hat{m}_f,\hat{\mu}),
~~~(\Delta, \omega_D)=\frac{e}{\kappa}(\hat{\Delta},\hat{\omega}_D)
\ee
where $\beta=\frac{e^4L^2}{\pi^2\kappa^2}$. The fluid densities $\hat{\rho},~\hat{n},~\hat{p}$ are 
  linearly proportional to the combination $\beta$ \cite{Hartnoll:2010gu}. The rescaling for $\lambda$ is chosen such that the dimensionless
  combination $\lambda\nu_0$ becomes $\lambda\nu_0
  =\hat{\lambda}\hat{\mu}\sqrt{\hat{\mu}^2-\hat{m}_f^2}/2$. Since we wish
    that $\delta \rho$ {\em etc.} scales the same way as $\rho$, the scaling
    for $\Delta$, and hence $\omega_D$ then follows. 
After this rescaling, the gap equation becomes 
$$\hat{\Delta}=2\hat{\omega}_D e^{-1/\hat{\lambda}\hat{\mu}\sqrt{\hat{\mu}^2-\hat{m}_f^2}}.$$
% To solve this system, w

We make the % following
standard homogeneous 
ansatz for the solution
\be\label{bggauan}
ds^2=L^2\big(-f(r)dt^2+g(r)dr^2+r^2(dx^2+dy^2)\big),~~~\hat{A}_t=h(r),
\ee
for which the equations of motion become
\bea\label{eomsystem}
\frac{1}{r}\bigg(\frac{f'}{f}+\frac{g'}{g}\bigg)-g(\hat{\rho}+\hat{p})&=&0,\nonumber\\
\frac{h'^2}{2f}+\frac{1}{r}\frac{f'}{f}+\frac{1}{r^2}-g(3+\hat{p})&=&0,\\
h''+h'\big(\frac{2}{r}-\frac{f'}{2f}-\frac{g'}{2g}\big)-\sqrt{f}g \hat{n}&=&0.\nonumber
\eea
Conservation of the energy-momentum tensor gives in addition:
\be\label{eomec}
(\hat{\rho}+\hat{p})f'-2\sqrt{f}\hat{n} h'+2f\hat{p}'=0.
\ee
The current is automatically conserved. 

Eqn. (\ref{eomec}) simplifies as 
\be
\frac{\hat{n}}{f} \bigg(\hat{\mu} f'-2\sqrt{f}h'+2f\hat{\mu}'\bigg)-\beta \frac{\hat{\Delta}^2}{\hat{\lambda}}\frac{\hat{\mu}'(2\hat{\mu}^2-\hat{m}_f^2)}{\hat{\mu}(\hat{\mu}^2-\hat{m}_f^2)}=0.
\ee
This equation can be integrated to give:
\be \hat{\mu}(r)=\frac{h}{\sqrt{f}}+\frac{1}{\sqrt{f}}\int_0 ^r d\tilde{r} \frac{\beta\hat{\Delta}^2 \sqrt{f}}{2\hat{n}\hat{\lambda}} \frac{\hat{\mu}'(2\hat{\mu}^2-\hat{m}_f^2)}{\hat{\mu}(\hat{\mu}^2-\hat{m}_f^2)} ,\ee
where the first term $h/\sqrt{f}$ is the leading order contribution and the second term is a sub-leading order contribution.
The position of the lower integration bound corresponds to the
integration constant. Since the prefactor of the integral, $1/\sqrt{f}$, is 
usually singular at the horizon, $r=0$, we chose the integration
constant to make sure that the integral itself vanishes at $r=0$.

The local value of the gap $\hat{\Delta}(r)$ is completely determined in terms of the local
chemical potential $\hat{\mu}$ and $\hat{\omega}_D$. The evolution of the local
chemical potential is completely determined by the equations of
motion, but the UV cut-off $\hat{\ome}_D$ requires additional
consideration. One option is to keep it constant. However, as $\hat{\mu}$ decreases, this would
rapidly invalidate our perturbative approach where
$\hat{\Delta}\ll\hat{\omega}_D\ll (\hat{\mu},\hat{\mu}-\hat{m}_f)$. We therefore use the freedom given to us
by the adiabatic approach to also promote it to slowly varying
parameter.
We choose to slave it to the chemical potential as
\be
\hat{\omega}_D=c\frac{\hat{\mu}^2-\hat{m}_f^2}{\hat{\mu}}.
\ee
For $c<1$ this ensures that our perturbative evaluation of the BCS
fluid holds.

\begin{figure}[h!]
 \begin{center}
% \begin{tabular}{cc}
% \includegraphics[width=0.5\textwidth]{zl1.pdf}
% \includegraphics[width=0.5\textwidth]{zl3.pdf}
% \end{tabular}
% \begin{tabular}{cc}
% \includegraphics[width=0.5\textwidth]{zl4.pdf}
% \includegraphics[width=0.5\textwidth]{zl2.pdf}
% \end{tabular}
\includegraphics[width=0.47\textwidth]{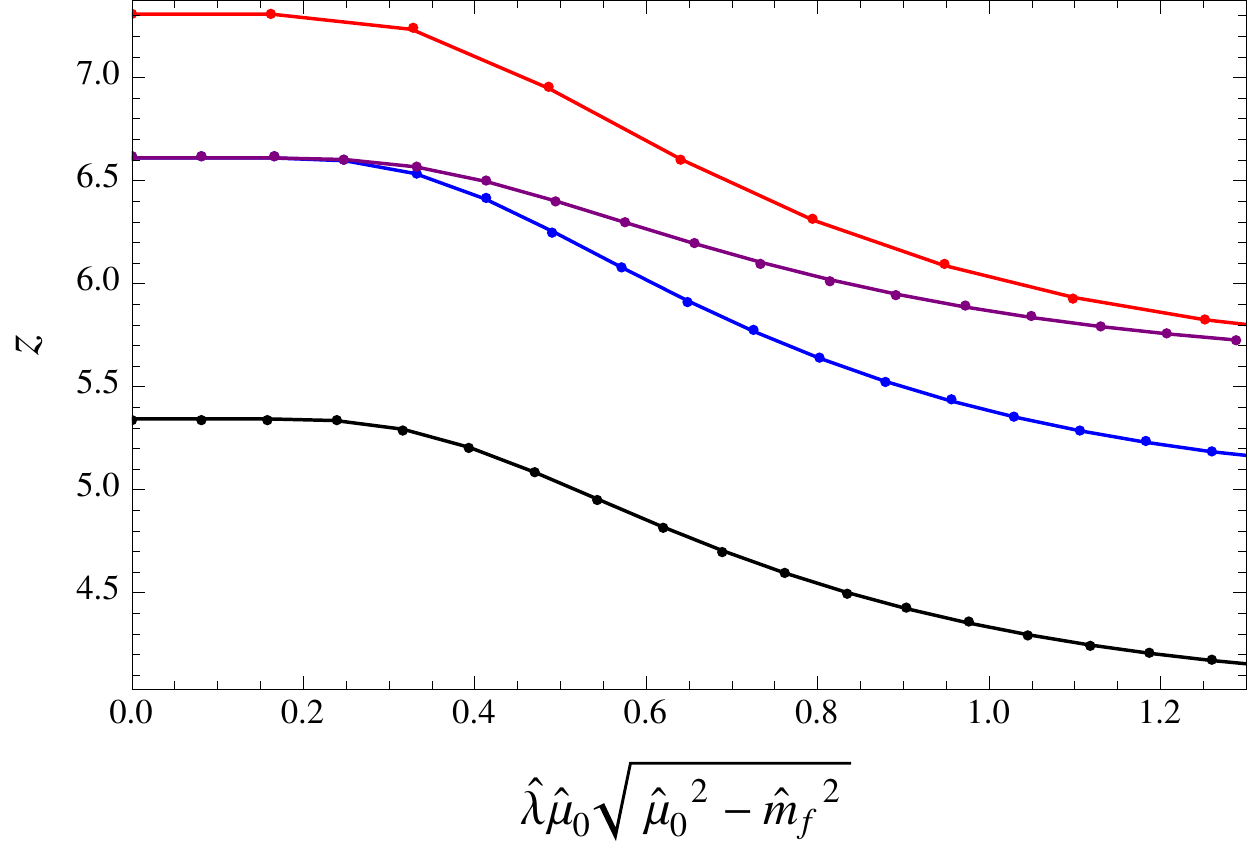}~~\includegraphics[width=0.52\textwidth]{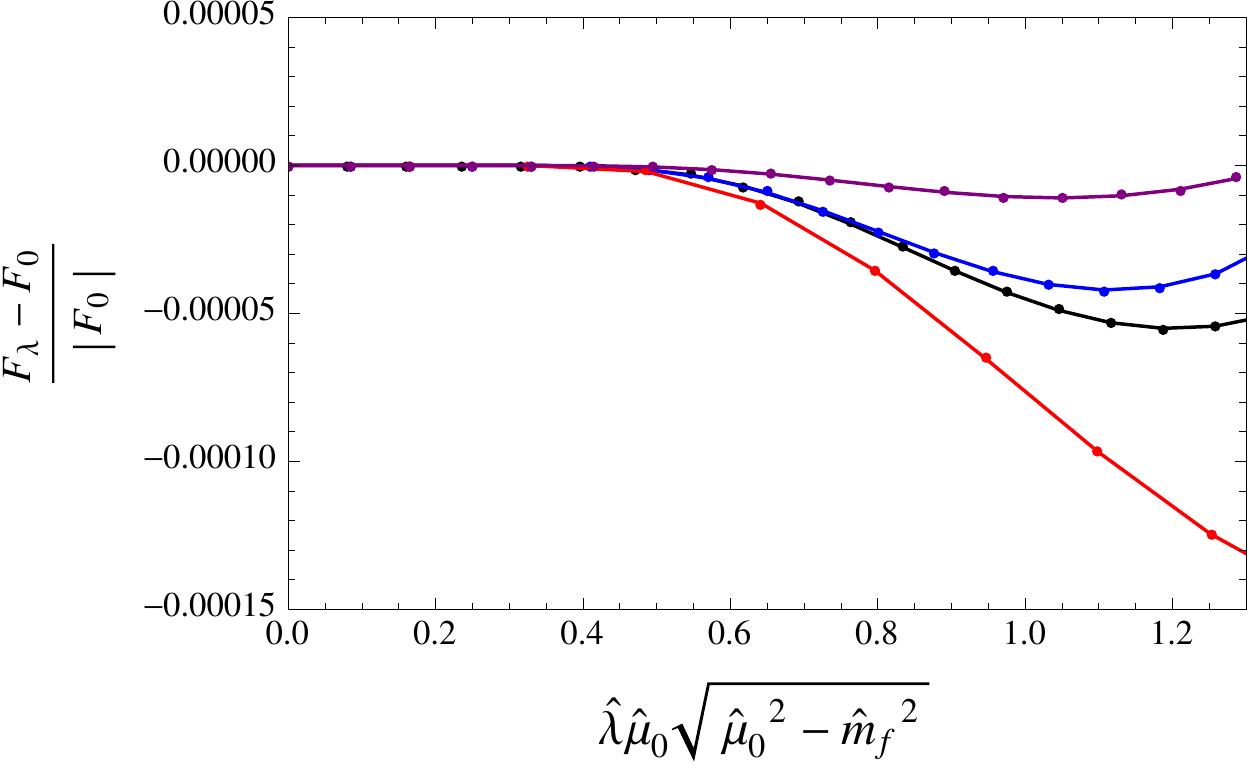}
\caption{\small The % dependence of the
  near horizon Lifshitz scaling exponent $z$ and the relative free
  energy $\big(F_{\hat{\lambda}}-F_{0}\big)/|F_0|$, with $F_0$ the free energy of the
  $\lambda=0$ electron star, as a function of the dimensionless coupling constant
  $\lambda \nu_0$ for different parameters:
  $\hat{m}_f=0.2,c=1/3,\beta=5~(\text{Blue})$, $\hat{m}_f= 0.3,c=1/3,\beta=5~ (\text{Red})$;  
$\hat{m}_f=0.2,c=1/4,\beta=5~(\text{Purple})$;
$\hat{m}_f=0.2,c=1/3,\beta=6~(\text{Black})$.
 For $\hat{\lambda}\hat{\mu}_0\sqrt{\hat{\mu}_0^2-\hat{m}_f^2} \lsim 1$ the free energy shows that the BCS
 star is the preferred groundstate.
The rising free energy beyond $\hat{\lambda}\hat{\mu}_0\sqrt{\hat{\mu}_0^2-\hat{m}_f^2} =1$ should not
be trusted. This is where perturbation theory
  breaks down.
}
\label{zlambda}
\end{center}
\end{figure}

We now follow the conventional procedure to find the solution. We search
for a scaling solution in the IR near the horizon where $r=0$, 
of the form
\bea 
\label{scalsol}
f=r^{2 z}~,~~
g=\frac{g_0}{r^2}~,~~
 h=h_0 r^z~,~~
 \hat{\mu}=\hat{\mu}_0.
\eea 
The scaling exponent is determined numerically (see
Fig. \ref{zlambda}).
We then perturb the solution
\bea f=r^{2 z} (1+f_1 r^{\alpha})~,~~
g=\frac{g_0}{r^2}(1+g_1 r^\alpha)~,~~
 h=h_0 r^z (1+h_1 r^{\alpha})~,~~
 \hat{\mu}=\hat{\mu}_0(1+\mu_1 r^{\alpha})
,~~\hspace{.25in}\eea
% \bea f&=&r^{2 z} (1+f_1 r^{\alpha}),\nonumber\\
% g&=&\frac{g_0}{r^2}(1+g_1 r^\alpha),\nonumber\\
%  h&=&h_0 r^z (1+h_1 r^{\alpha}), \nonumber\\
%  \hat{\mu}&=&\hat{\mu}_0(1+\mu_1 r^{\alpha}),\eea 
and search for a perturbation where the coefficient $f_1$ can remain a
free parameter. There are multiple such solutions and we seek the one
with positive exponent $\alpha>0$. This corresponds to a perturbation
of the IR by an irrelevant operator and we can integrate this flow up
to an asymptotically AdS$_4$ solution. The exponent $\alpha$ is also
determined numerically.
%\comment{DO WE KNOW $\alp$ ANALYTICALLY?}
% where $z$ is the Lifshitz scaling exponent and $\alpha>0$ is a positive constant. (See Fig. \ref{zlambda}.) The parameters $h_0$, $g_0$, $\mu_0$,  z and $\alpha$ can all be determined by the equations of motion at the horizon. Then the perturbations of order $r^{\alpha}$ can drive the system to asymptotic AdS$_4$ boundary.
%We can solve from the equation above that \be \mu_0=h_0\ee and \be \mu_1=\frac{(h_1-f_1/2)}{1-\frac{3c^2\alpha (2h_0^2-m^2) e^{-\frac{4}{\lambda h_0 \sqrt{h_0^2-m^2}}}}{\lambda \beta h_0^3 \sqrt{h_0^2-m^2} (\alpha+z)}}\ee

When integrating this system numerically from the horizon to the
boundary one encouters the star edge $r_s$, which is determined by 
\be
\hat{\mu}(r_s)=\hat{m}_f.
\ee
At this point all fluid densities vanish.
Outside the star, the geometry is described by RN black hole with the metric  
\be
f=c^2\bigg(r^2-\frac{M}{r}+\frac{Q^2}{2r^2}\bigg),~~~
g=\frac{c^2}{f},~~~
h=c(\mu-\frac{Q}{r}).
\ee 
The charge $Q=Q_{\text{tot}}.$ is the total charge contained within
the interior of the star.

\begin{figure}[h!]
\begin{center}
\begin{tabular}{cc}
\includegraphics[width=0.45\textwidth]{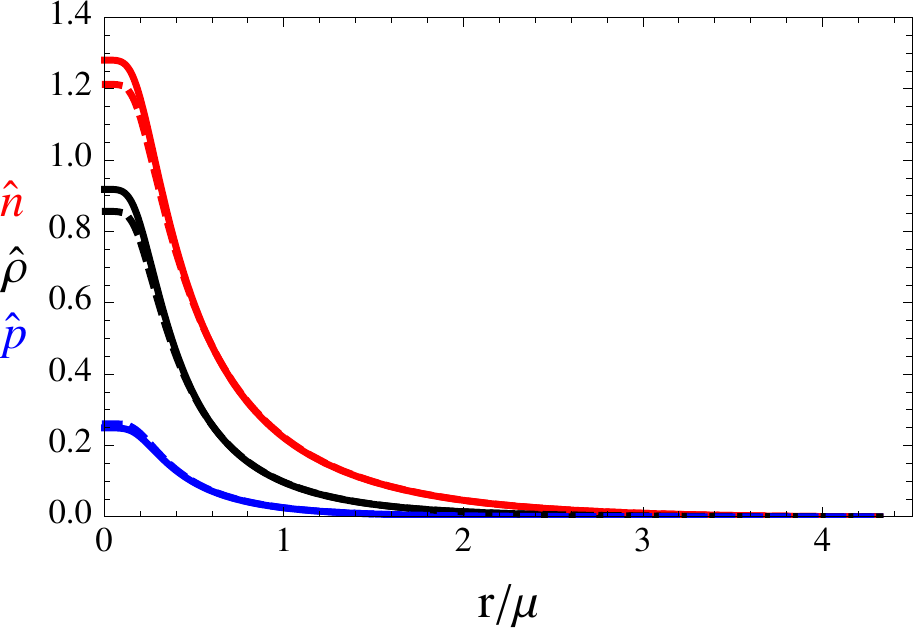}
\includegraphics[width=0.47\textwidth]{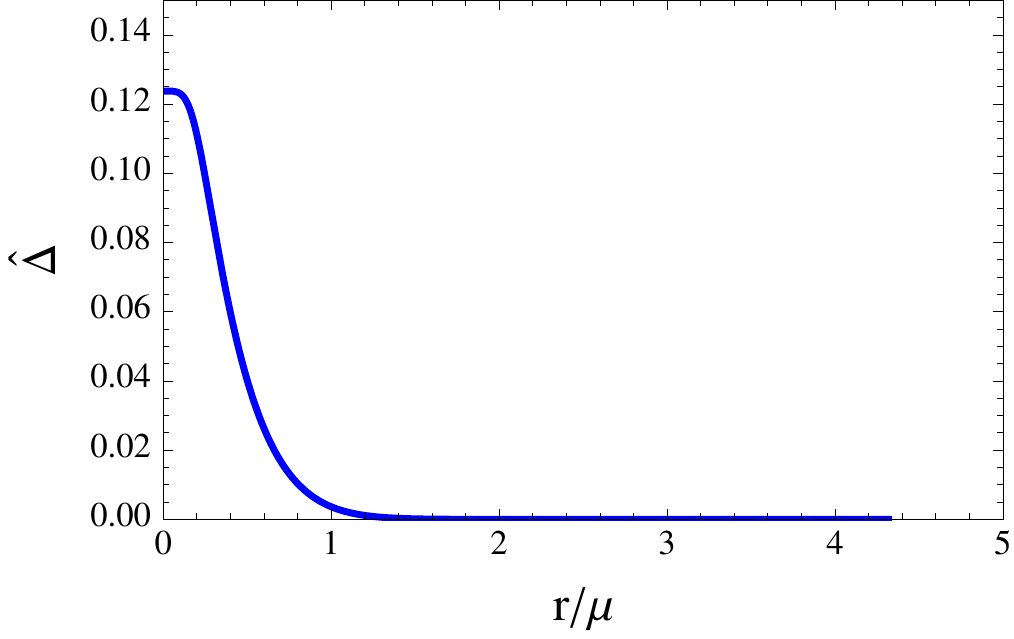}
\end{tabular}
\caption{\small The BCS star profile as a function of the radial coordinate for $\hat{m}_f=0.2, c=1/3, \beta=5$ and
  $\hat{\lambda}\hat{\mu}\sqrt{\hat{\mu}^2-\hat{m}_f^2}=$ 0.649. 
  Left: from top to bottom the fluid densities 
  $\hat{n}_f$,$\hat{\rho}_f$,$\hat{p}_f$ of the BCS star
    (solid line) compared to the
    electron star
  (with same $\hat{m}_f, \beta$ and $ \hat{\lambda}\hat{\mu}\sqrt{\hat{\mu}^2-\hat{m}_f^2}=0$;  dashed line).
Both the charge(number) and energy density increase compared to the electron star. The star edges $r_s/\mu$ for ES and BCS are 4.320 and 4.329 respectively. 
  Right: the order parameter $\hat{\Delta}$ in the BCS star solution.
%\comment{What is the value of $r$ at the edge of the star?}
\\
}
\label{fluid}
\end{center}
\end{figure}

The total solution is characterized by four dimensionless parameters
$\hat{m}_f,\beta,c,\hat{\lambda}\hat{\nu}_{0}$. Here we use the local density
$\hat{\nu_{0}}\equiv\hat{\mu}\sqrt{\hat{\mu}^2-\hat{m}_f^2}|_{r=0}$ at the
horizon, Eq. (\ref{scalsol}) to make the BCS coupling dimensionless. Fig. \ref{fluid} shows for one such
solution both the behavior of the fluid and the condensate in the
fluid region. The densities of the fluid are cleanly decreasing along the radial coordinate. % which means that there is always only one outer edge.
Our interest here is the transition to pairing and condensation. This
is controlled by the dimensionless BCS coupling $\lambda\nu_0$ and we
study the system as this is varied.
In Fig. \ref{zlambda}, we show the dependence of the near horizon
scaling exponent $z$ on $\hat{\lambda}\hat{\nu}_0$ for various values of $\hat{m}_f, c, \beta$.

The relative value of the free energy of BCS star backgrounds
w.r.t. the free energy at $\lambda=0$ is shown as a function of the
coupling constant
$\lambda\nu_0=\hat{\lambda}\hat{\mu}\sqrt{\hat{\mu}^2-\hat{m}_f^2}/2$
in Fig \ref{zlambda}. The free energy can be determined from the parameters
of the exterior solution
\be
F/\mu^3=(M-\mu Q)/\mu^3.
\ee 
The free energy at $\hat{\lambda}=0$ is the free energy of the
electron star \cite{Hartnoll:2010gu}. As $\hat{\lambda}$ goes larger,
the free energy decreases. This shows that in the bulk, BCS star is a
more stable solution due to the local attractive interactions between
fermions. Note that when $\hat{\lambda}\hat{\nu}_0$ approaches order 1, the free
energy starts to grow again. We have found that it does so in all
cases for some $\hat{\lambda}\hat{\nu}_0 \gsim 1$. However, this is the regime
where perturbation theory fails, and the computation is not reliable
for these large values.

%%%%%%%%%%%%%%%%%%%%%%%%%%%%%%%%%%
%%%%%%%%%%%%%%%%%%%%%%%%%%%%%%%%%%

%%%%%%%%%%%%%%%%%%%%%%%%%%%%%%%%%%
%%%%%%%%%%%%%%%%%%%%%%%%%%%%%%%%%%

% %%%%%%%%%%%%%%%%%%%%%%%%%%%%%%%%%%%%%%%%%
% \subsection{Free energy}
% %%%%%%%%%%%%%%%%%%%%%%%%%%%%%%%%%%%%%%%%%
% We can integrate this system numerically from the horizon to the boundary. There is an edge of the BCS star $r_s$, which is determined by 
% \be
% \hat{\mu}(r_s)=\hat{m}_f.
% \ee

% Outside the star, there is no fluid excitations and the geometry is described by RN black hole with the metric  
% \be
% f=c^2\bigg(r^2-\frac{M}{r}+\frac{Q^2}{2r^2}\bigg),~~~
% g=\frac{c^2}{f},~~~
% h=c(\mu-\frac{Q}{r}).
% \ee 
% We have $Q=Q_{\text{tot}}.$

%%%%%%%%%%%%%%%%%%%%%%%%%%%%%%%%%%
% \begin{figure}[h!]
% \begin{center}
% % \begin{tabular}{cc}
% % \includegraphics[width=0.5\textwidth]{fl1.pdf}
% % \includegraphics[width=0.5\textwidth]{fl3.pdf}
% % \end{tabular}
% % \begin{tabular}{cc}
% % \includegraphics[width=0.5\textwidth]{fl4.pdf}
% % \includegraphics[width=0.5\textwidth]{fl2.pdf}
% % \end{tabular}
% \includegraphics[width=0.7\textwidth]{FData.pdf}
% \caption{\small Free energies of the BCS backgrounds as a function of the coupling constant $\lambda$ for $\hat{m}_f=0.2 (\text{Blue}), 0.3 (\text{Red}),
%   c=1/3,\beta=5$; $\hat{m}_f=0.2,c=1/4,\beta=5 (\text{Orange})$;
%   $\hat{m}_f=0.2,c=1/3,\beta=6 (\text{Black})$. Perturbation theory
%   breaks down at large $\lambda$.
% %\comment{IMPROVE CAPTION}
% }
% \comment{AGAIN THE SAME THREE PLOTS WITH VARYING $c$ AS FOR $z$}
% \label{freeen}
% \end{center}
% \end{figure}
%%%%%%%%%%%%%%%%%%%%%%%%%%%%%%%%%%

%%%%%%%%%%%%%%%%%%%%%%%%%%%%%%%%
%%%%%%%%%%%%%%%%%%%%%%%%%%%%%%%%
\section{Properties of the dual field theory: evidence of superconductivity}\label{sec3}
%%%%%%%%%%%%%%%%%%%%%%%%%%%%%%%%
%%%%%%%%%%%%%%%%%%%%%%%%%%%%%%%%

In the last section we showed that our BCS star is more stable than
the electron star solution at zero temperature and nonzero $\lambda$
and it can be seen as a continuous interaction driven quantum phase
transition at $T=0$. In this section we will show the
evidence that this BCS star corresponds to a superconducting state at
the boundary. We cannot show this by conventional
  holographic means. Due to the fact that no collective fields extend
  beyond edge of the star --- an artifact of the Thomas-Fermi
  approximation --- there is no leading coefficient to be read off
  near the AdS boundary. Instead we will first show that there is a
gap in the dual Fermi spectral function which resembles that of a
superconducting state. % Then we will show that the Luttinger's theorem
% at the boundary is violated. The violation indicates that charge
% has disappeared from the Fermi-liquid.
Next we will study the change in the constituent charge
  densities, and show explicitly that charge disappears from the Fermi
  liquid into the bosonic sector. 
This shows that Cooper
pairs have formed and have carried away the charge. Finally we compute
the conductivity at small frequency and show that it has the hallmark
characteristics of a holographic superconductor: a delta-function peak
at zero frequency (foremost a consequence of momentum conservation) and a soft gap at $\omega<\Delta$.

%%%%%%%%%%%%%%%%%%%%%%%%%%%%%%%%%%%%%%
\subsection{Gap in the Fermi spectral function}
%%%%%%%%%%%%%%%%%%%%%%%%%%%%%%%%%%%%%%

To calculate the dual Fermi spectral function, we need to consider Fermi perturbations in the bulk which couple to the local gap function $\Delta$ with a BCS interaction as follows:
\be
S_{\text{probe}}=\int d^4x \sqrt{-g}\bigg[-i \bar\Psi(\Gamma^\mu  \mathcal{D}_\mu-m_f)\Psi +\frac{1}{2}\Delta^* \bar\Psi_c\Gamma^5\Psi-\frac{1}{2}\Delta\bar\Psi \Gamma^5\Psi_c\bigg].
\ee
The probe fermion has the same mass and charge as the fermion that
constitute the bulk star solution before the scaling. The
  scaling, however, does not act uniformly on the probes
  \cite{Cubrovic:2011xm}. After the scaling, an explicit dependence on
  the ratio $L/\kap$ remains. This is the reflection of the inherent
  quantum mechanical nature of fermions. We will not consider this in
detail because these coefficients are not important for showing the physical results related to the gap.

The BCS interaction term couples two modes of opposite spin, which have the same spectrum.\footnote{The eigenstates of the Dirac equation have either a left-pointing spin or right pointing spin w.r.t. the momentum with independent Fermi surfaces. Due to a spin-orbit-like coupling with the background electric fields \cite{Herzog:2012kx,{Alexandrov:2012xe}, inprogress2}, these Fermi surfaces are slightly split $k_{F_L}\neq k_{F_R}$. Despite this split, a  spin-zero BCS pairing at $k=0$ is still allowed as the left-pointing spin at $k_{F_L}$ w.r.t. the momentum points in the opposite direction as the left-pointing spin at $-k_{F_L}$; and similarly for $k_{F_R}$. In the fluid limit here, this detail is not directly apparent, as it gets subsumed in the many different Fermi surfaces corresponding to each radial mode of the Dirac field.
\bigskip
\par
}
%{\commentr{There is in fact a tiny splitting in
%    the two Fermi surfaces due to the presence of the background
%    electric field in AdS \cite{Herzog:2012kx,{Alexandrov:2012xe},inprogress2}. 
%In the fluid limit here, this
%    splitting is so small, it is not relevant.}} 
The gap in the fermion spectrum is simply the level repulsion from coupling two
  degenerate states. % , and this coupling is crucial for generating
                    % gapped spectrum in the Fermi spectral function.
The Dirac equation with BCS interaction is \be i(\Gamma^{\mu} D_{\mu}-m_f)\Psi+
\Delta \Gamma^5\Psi_c=0.
\ee
After rescaling \be\psi= (-g g^{rr})^{1/4}\Psi,\ee we have 
\be\label{eomdiracbcs}
(\Gamma ^r \partial _r+\Gamma^{\mu}k_{\mu}-m_f)\psi (r,k,\omega)-\Delta C\Gamma^5\Gamma^0\psi^*(r,-k,-\omega)=0
\ee 
in the momentum space. 
Using \be \psi=(\psi_1,\psi_2)^T,\ee
equation (\ref{eomdiracbcs}) can be written as 
\be\big(-\sqrt{g^{rr}}\sigma^3\partial_r \mp i\sqrt{g^{xx}}\sigma^2 k+(\omega+ A_t)\sqrt{g^{tt}}\sigma^1-m_f\big)\psi_{1,2}(r,k,\omega)\pm i\Delta\sigma^1\psi_{2,1}^*(r,-k,-\omega)=0\ee
from which we observe 
$\psi_1(r,k,\omega)$ is coupled to $\psi_2^*(r,-k,-\omega)$ and $\psi_2(r,k,\omega)$
is coupled to $\psi_1^*(r,-k,-\omega)$. From the free Dirac equation of
motion we can see that the spectrum of $\psi_1(r,k,\omega)$ and
$\psi_2^*(r,-k,-\omega)$ are the same at $\omega=0$.
This is the degenerate point where the BCS interaction couples causes a gap.

% At the horizon, $\Delta$ is finite, so the interaction term is sub-leading compared to other terms.
To calculate the dual Green's function, we should first specify the
near horizon boundary conditions for this system. Following
\cite{Faulkner:2009am,Liu:2012tr}, we % work in the perturbation limit
% where
treat the BCS coupling term % is treated
as a perturbation. 
This is consistent since both at the horizon and at the boundary,
$\Delta$ is finite, so the  interaction term is sub-leading compared
to other terms. 
At the horizon we must choose infalling boundary conditions to
obtain the retarded Green's function in the dual boundary
theory. They can be chosen independently for  $\psi_1(r,k,\omega)$ and
$\psi_2^*(r,-k,-\omega)$. To solve the system, we can chose as a basis the
linearly independent choice I where $\psi_1(r,k,\omega)=0$,
$\psi_2^*(r,-k,-\omega)$ is ingoing and choice II where $\psi_1(r,k,\omega)$ is
ingoing while $\psi_2(r,-k,-\omega)=0$. Solving the Dirac equation with
these two independent horizon boundary conditions, we obtain two sets
of values at the AdS boundary at $r=\infty$. % At $r=\infty$ the
                                % coupling term again becomes
                                % sub-dominant and can be ignored.
As the BCS coupling is again subleading, the general form of
the boundary behavior is  
\be\psi_1^{\text{I,II}}(k,\omega)=A_1^{\text{I,II}} r^m \begin{pmatrix} 0\\ 1  \end{pmatrix}
 +B_1^{\text{I,II}}r^{-m} \begin{pmatrix} 1\\ 0  \end{pmatrix}
 \ee 
 and \be \psi_2^{* \text{I,II}}(-k,-\omega)=A_2^{* \text{I,II}} r^m \begin{pmatrix} 0\\ 1  \end{pmatrix}
 +B_2^{* \text{I,II}}r^{-m} \begin{pmatrix} 1\\ 0  \end{pmatrix}.
 \ee 
where the superscript I,II refers to the choice of horizon boundary
conditions.

We therefore obtain a matrix of responses $B$ to the various
  sources $A$, % i.e. the Green's function is defined as
\be\begin{pmatrix} 
B_1^\text{I}  & B_1^{\text{II}}\\ 
B_2^{* \text{I}} & B_2^{* \text{II}}  
\end{pmatrix}=\begin{pmatrix} G_{O_1O_1^\dagger} & G_{O_1O_2} \\ G_{O_2^\dagger O_1^\dagger} &G_{O_2^\dagger O_2} \end{pmatrix}
\begin{pmatrix} 
A_1^\text{I}  & A_1^{\text{II}}\\ 
-A_2^{* \text{I}} & -A_2^{* \text{II}}  
\end{pmatrix}.
\ee
 The Green's function can then be calculated as $G= B A^{-1}.$ 
%Here we work in the weak potential limit. % In the first boundary
                                % condition, $\psi_2^*(r,-k,-\omega)$ is of
                                % order $\mathcal{O}(1)$ with order
                                % $\lambda^2$ corrections and
                                % $\psi_1(r,k,\omega)=0$ is of order
                                % $\lambda$. In the second boundary
                                % condition, it is similar. Then we
                                % have the off-diagonal elements in A
                                % and B of order $\lambda$ and the
                                % diagonal elements of order
                                % $\mathcal{O}(1)
                                % +\mathcal{O}(\lambda^2)$.
In the absence of a BCS interaction $G$ is diagonal. In the
perturbative limit we use here, the off-diagonal terms are of order
$\Delta$ and the diagonal terms receive corrections of order $\Delta^2$.

% Before introducing
In the absence of 
the BCS interaction, the system has poles at $\omega=0$ and we can define
the (set of) Fermi momentum(momenta) $k_F$ as the value(s)
where the leading fall-off of the (diagonal) solution vanishes
$A_1^{\text{I}}(k_F,0)=0$ and $A_{2}^{*\text{II}}(-k_F,0)=0$. For a star solution which exists in the WKB limit,
there are usually multiple Fermi surfaces
\cite{Hartnoll:2011dm,Iqbal:2011in,Cubrovic:2011xm}. Here we take
$k_F$ to be the largest % one which corresponds to the primary
%such 
Fermi surface --- the primary Fermi surface --- though the following arguments apply to any of the Fermi surfaces.

Including now the BCS interaction, the source matrix $A$ near $(k=k_F,\omega=0)$ is % the diagonal elements of the solution $A(k_F,\omega)$ at the Fermi momentum behaves % as $A(k_F,\omega)\sim \mathcal{O}(\omega)+\mathcal{O}(\omega^2,\lambda^2)$ and the off-diagonal ones $A(k_F,\omega)\sim \mathcal{O}(\lambda)$. We can write $A$ as
\be 
A(k_F,\omega) \sim \begin{pmatrix} a_1^\text{I} \omega &
  a_1^{\text{II}}\Delta \\ -a_2^{*\text{I}} \Delta &
  -a_2^{*\text{II}} \omega \end{pmatrix} + \cO(\Delta^2,\omega^2)
\ee 
at the leading order, where $a_1^{\text{I,II}}$ and $a_2^{*\text{I,II}}$ are % UV
constants  % which are usually
of order $\mathcal{O}(1)$ near the Fermi surface. From this expression
we can already see that there is a gap at the Fermi surface % where
                                % the gap
with size
$\Delta$. In \cite{Faulkner:2009am}, these coefficients % were written
                                % out more
are obtained
explicitly by expanding the system near $\omega=0$ and at
$k=k_F$. Denoting the (normalizable) solution to the Dirac equations for which
$A(k,\omega)$ vanishes at
$\omega=0$ and $k=k_F$ as $\psi_{1}^{\text{I}}(k_F,0)=\xi_1^{(0)}$ and
$\psi_2^{*\text{II}}(-k_F,0)=\xi_2^{(0)}$, they find 
\cite{Faulkner:2009am}:
 \be G_R^{-1} (k_F,\omega)\sim \begin{pmatrix} \omega P_{1} & Q_1\\Q_2 &
  \omega P_2 \end{pmatrix},\ee where 
 \bea P_{\alpha}&=&\int dr
 \sqrt{g_{rr}}\bar{\xi}_{\alpha}^{(0)}\sqrt{g^{tt}}\xi_{\alpha}^{(0)}(-1)^{\alpha},
 \nonumber\\\label{pqvalue} 
Q_{1}&=&\int dr \sqrt{g_{rr}}\bar{\xi_{1}}^{(0)}  i \Delta
\xi_{2}^{(0)}, \\  
Q_{2}&=&\int dr \sqrt{g_{rr}}\bar{\xi_{2}}^{(0)}  i \Delta \xi_{1}^{(0)}.\nonumber\eea
%These give explicit calculations to the coefficients in the matrix $A$.
Diagonalizing one finds a gap for
%The gap is then determined by 
\be\label{gap} 
|\omega| < \sqrt{Q_1Q_2/P_1P_2}
\ee
which is of order $\Delta$ taking value at the horizon. 
% Thus we can see that the dual field theory indeed has a gap an
This gap in the fermion spectral function indicates that the
  field theory should be in a superconducting state. Similar to the holographic lattice gap \cite{Liu:2012tr}, this gap is only a pseudo-gap in the sense that the $G^{-1}_R$ is only zero at one special $\omega$ and away from that frequency there will be small spectral weights.

%\comment{Multiple FS means multiple gaps??}

%%%%%%%%%%%%%%%%%%%%%%%%%%%%%%%%%%%%%%%%%%%%%%
\subsection{
Superconductivity induced changes in the charge density}
%%%%%%%%%%%%%%%%%%%%%%%%%%%%%%%%%%%%%%%%%%%%%%

The gap in the spectral function of the dual CFT on the
  boundary is the consequence of the superconducting core in the BCS
  star, even though its wavefunction does not extend to boundary. It
  is readily understood why: the lifting of the degeneracy need
  only to happen at one point in the interior. 
% This boundary gap can be seen as a boundary effect of this BCS star. The gap of the boundary is related to the integration of the gap (\ref{pqvalue}, \ref{gap}) in the bulk.
Another effect that persists into the dual CFT is the
redistribution of the 
charge density of the system.  The boundary charge density % comes
arises 
from the boundary value of the Maxwell field and when there is no
contribution of charge density % within
from inside 
the horizon, the boundary charge density is also equal to the
integration of the bulk charge density along the radial direction
\cite{Iqbal:2011bf,Hartnoll:2011dm}. 
Assuming that all fermions in region II immediately pair up at any finite $\lambda$, 
we can separate the total boundary charge density into two parts: the
free charge density $Q_{\text{free}}$ from the fermions in region I, and the charge density which corresponds to paired fermions $Q_{\text{pair}}=Q_{\text{total}}-Q_{\text{free}}$ in the bulk. They can be obtained by the bulk integration of the charge density as follows
\be\label{QQinSecII}
Q_{\text{free}}=\int_0^{r_s} dr r^2\sqrt{g_{rr}} n_{\text{I}}^{\text{FL}},~~~~ Q_{\text{total}}=\int_0^{r_s} dr r^2\sqrt{g_{rr}} n,
\ee 
with $n_{\text{I}}^{\text{FL}}$ in (\ref{freefermionsecII}).

A further quantity of interest is the deviation from the exact
equation of state of the free Fermi liquid. This is qualitatively
captured by the amount of charge in the deviation density 
\be Q_{\text{dev}}=
\int_0^{r_s} dr r^2 \sqrt{g_{rr}} \delta n_{\text{total}}\ee 
with  $\delta
n_{\text{total}}$ in (\ref{deltan}).

\begin{figure}[h!]
\begin{center}
\begin{tabular}{cc}
\includegraphics[width=0.47\textwidth]{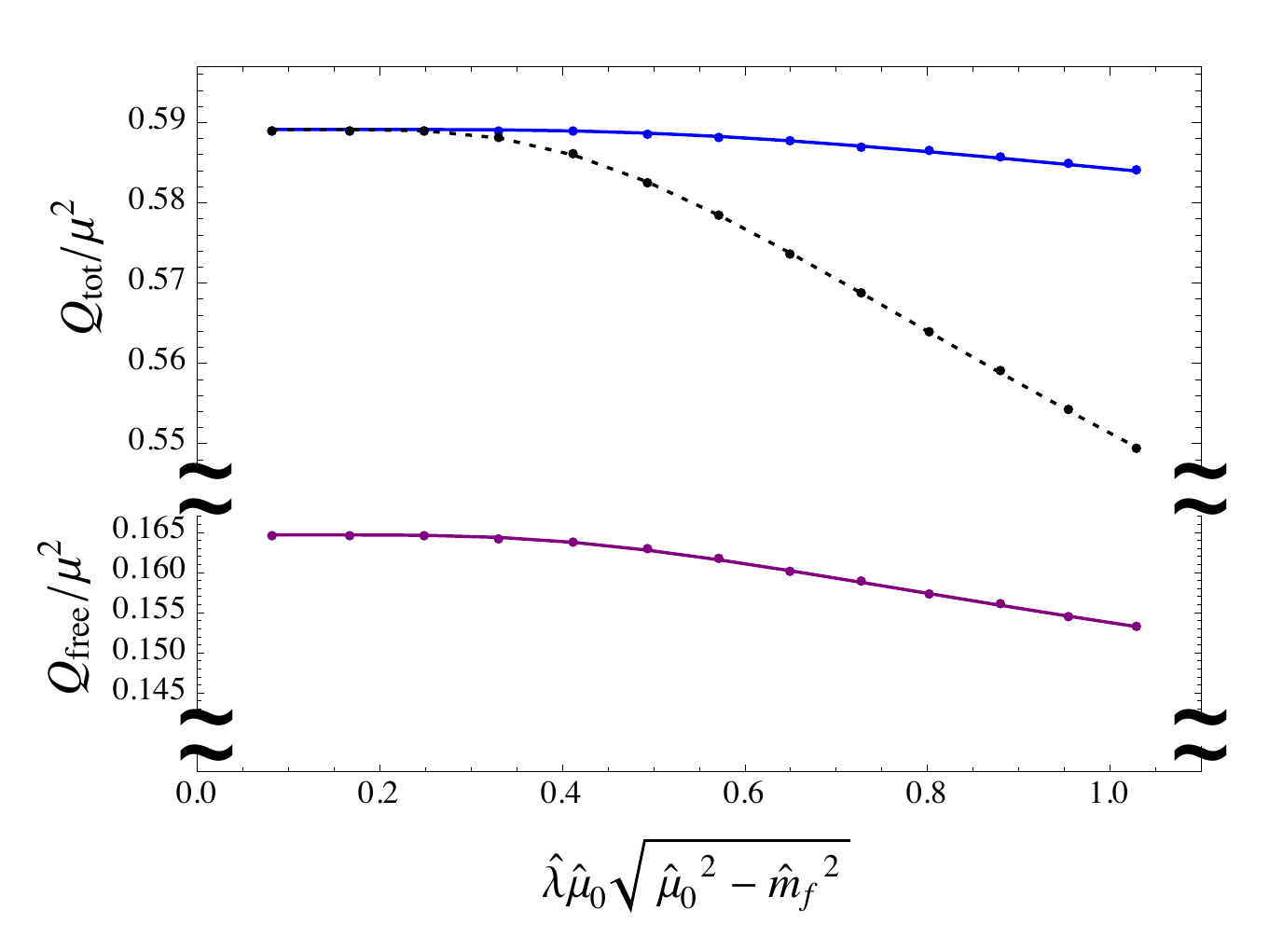}~~
\raisebox{.04in}{\includegraphics[width=0.47\textwidth]{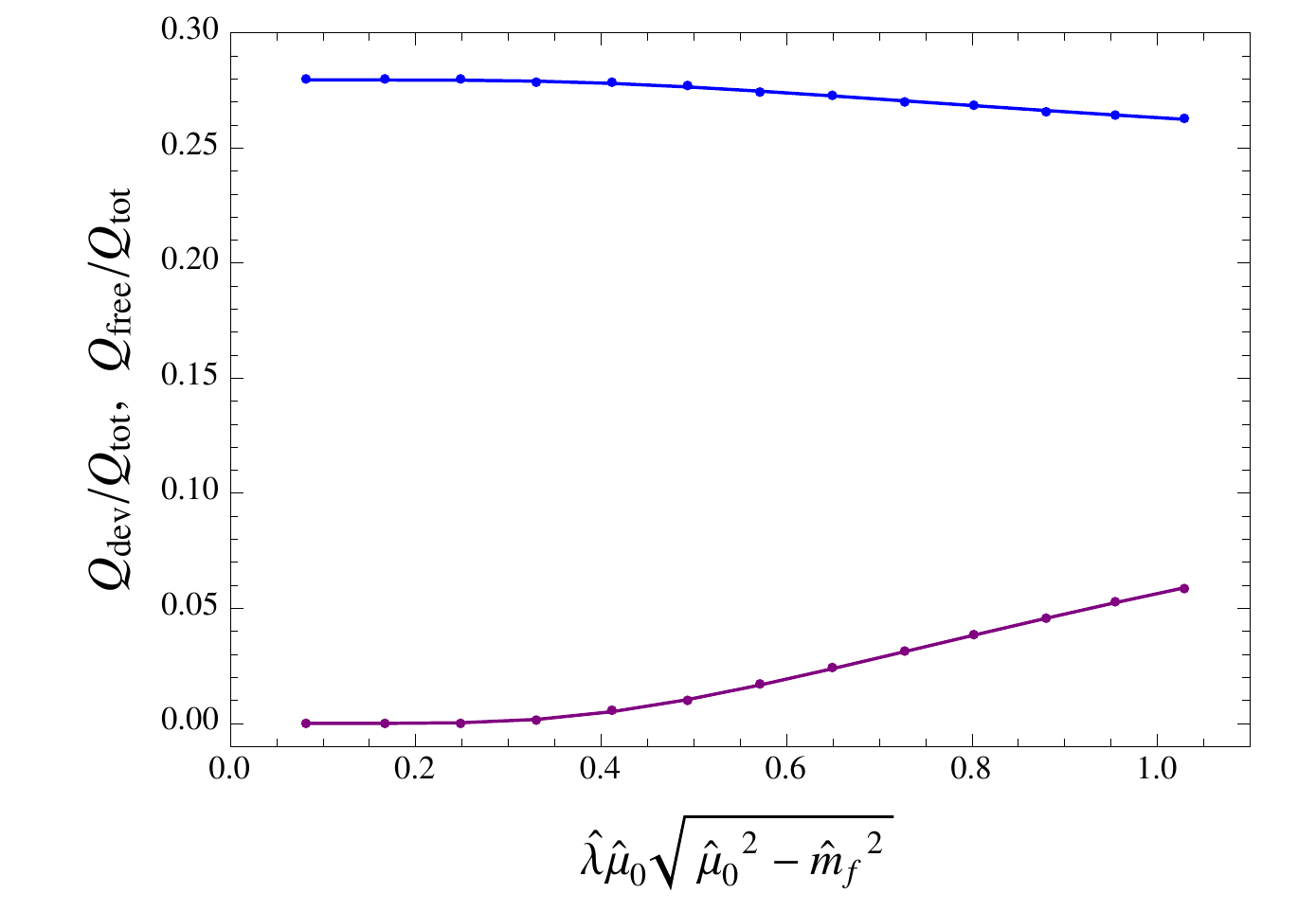}}
\end{tabular}
\caption{\small The total and free fermion charge density,
  $Q_{\text{total}}$ and $Q_{\text{free}}$ as a function of
  $\hat{\lambda}\hat{\mu}_0\sqrt{\hat{\mu}_0^2-\hat{m}_f^2}$ for
  $\hat{m}_f=0.2, c=1/3, \beta=5$. The %region between the 
  dashed (black) line in the left figure between $Q_{\text{free}}$
  and $Q_{\text{total}}$ shows the effect of the change of the
  equation of state compared to the standard free Fermi liquid: it is the
  contribution %$\delta Q_{\text{dev}} = \int \delta n$
  $Q_{\text{dev}}$. The
  decrease in the free fermion contribution is compensated by the
  change in the equation of state. but only partially. Note that
  $Q_{\text{total}}$ {\em decreases} as a function of the coupling
  $\lambda\nu_0$ indicating that it becomes progressively more
  difficult to excite charged carriers as the BCS coupling is turned
  on.
On the right
  hand side we show the relative contributions
  $Q_{\text{free}}/Q_{\text{total}}$,
  $Q_{\text{dev}}/Q_{\text{total}}$.
This visibly shows the pairing taking place as the deviation from the free Fermi liquid equation of state grows.
}
\label{ratio}
\end{center}
\end{figure}

In Fig. \ref{ratio}, we show both the absolute and relative values of
these charge density contributions compared to the total charge $Q_{\text{total}}$ as a function of the BCS coupling
$\lambda$. Perhaps counterintuitively, the total charge density
$Q_{\text{total}}$ (in units of the chemical potential $\mu$) {\em
  decreases} as we increase the BCS coupling
$\hat{\lambda}\hat{\nu}_0$. 
%That the charge density in BCS theory changes for fixed
%chemical potential is a known effect in perturbative BCS that
%underlies the charging of vortices
%\cite{Feiner,Khomskii}. 
It is known in condensed matter physics that the charge density is generically influenced by the condensate when the normal state is not invariant under charge conjugation on the scale of the superconducting gap. In weak coupling BCS it can be calculated that the charge density changes with a difference proportional to the order of the gap, but because it is weakly coupled the gap is small enough for this difference to be ignored. However, when the superconductor gets more strongly coupled such that the density of states is asymmetric around the Fermi surface on the scale of the gap the charge density (or either the chemical potential in the case of the grand canonical ensemble) changes when the order parameter develops. A typical example of the consequences of this very basic property is that vortices (where the core turns normal) are charged in more strongly coupled superconductors as confirmed by experiments in high Tc superconductors \cite{Feiner,Khomskii}. 
%In perturbative BCS the
%  charge density increases. 
Here in our holographic model the
decrease shows that the interaction makes charged excitations more
difficult to populate rather than easier.
Part of this decrease is simply due to Bose-Fermi competition: we see this as
the decreasing 
contribution from the free fermions in region I. Condensing Cooper
pairs do compensate this decrease, but not sufficiently so to
increase $Q_{\text{total}}$. The fact that Cooper pairs do form is
shown by the non-vanishing deviation from the Fermi liquid equation of
state $Q_{\text{dev}}$. 
The non-vanishing charge density in the Cooper-pair sector
  shows explicitly that the dual ground state is charged and
 breaks the $U(1)$ gauge symmetry.

Note that our definition of $Q_{\text{free}}$ only
counts the fermions in region I. Therefore it does not equal
$Q_{\text{total}}=Q_{\text{I}}+Q_{\text{II}}$ at $\lambda=0$.

%%%%%%%%%%%%%%%%%%%%%%%%%%%%%%%%%%
\subsection{Conductivity at low frequency}
%%%%%%%%%%%%%%%%%%%%%%%%%%%%%%%%

For completeness, we also consider the behavior of conductivity at low frequency for the dual field theory.  Following \cite{Hartnoll:2010gu},
we consider the time dependent perturbations:
\be
A_x=\frac{eL}{\kappa}\delta a_x(r) e^{-i\omega t},~~
g_{tx}=L^2\delta g_{tx}(r) e^{-i\omega t},~~
u_x=L\delta u_x(r) e^{-i\omega t}. 
\ee

The equations of motion for these fluctuations are 
\bea
\hat{n}\delta a_x+(\hat{\rho}+\hat{p})\delta u_x&=&0,\nonumber\\ \label{eomfluctuations}
\delta g_{tx}'-\frac{2}{r}\delta g_{tx}+2h'\delta a_x&=&0,\\
\delta a''_x+\frac{1}{2}\bigg(\frac{f'}{f}-\frac{g'}{g}\bigg)\delta a'_x+\omega^2\frac{g}{f}\delta a_x+\frac{h'}{f}\bigg(\delta g'_{tx}-\frac{2}{r}\delta g_{tx}\bigg)+g \hat{n} \delta u_x&=&0.\nonumber
\eea
Substituting the first and second equations in (\ref{eomfluctuations}) into the third, we obtain the EOM for $\delta a_x$
\be\label{eomax}
\delta a''_x+\frac{1}{2}\bigg(\frac{f'}{f}-\frac{g'}{g}\bigg)\delta a'_x+\bigg(\frac{g\omega^2}{f}-\frac{2h'^2}{f}-\frac{g\hat{n}^2}{\hat{\rho}+\hat{p}}\bigg)\delta a_x=0.
\ee
The near horizon geometry of the BCS star is a Lifshitz
geometry controlled by the dynamical critical exponent $z$ and 
the solution of  (\ref{eomax}) in the Lifshitz region with the 
infalling boundary condition 
$\delta a_x^{(\text{in})}\sim e^{i\frac{\omega \sqrt{g_0}}{z r^z}}$
is
\be\label{nhdeltaax}
\delta a_x=\sqrt{\omega}r^{-\frac{z}{2}}H^{(1)}_\frac{\sqrt{4c_0+z^2}}{2z}\bigg[\frac{\sqrt{g_0}\omega r^{-z}}{z}\bigg]. 
\ee
Here the freedom to set the
  amplitude in the fluctuation equation is used to set it proportional to $\sqrt{\omega}$ --- this way the
  leading order
  coefficient for the near-horizon infalling wave does not depend on frequency, $H^{(1)}_{\nu}(x)$ is the Hankel function of first kind and the constant $c_0$
equals 
\be c_0=2h_0^2
z^2+\frac{g_0\beta}{h_0}\bigg(\frac{1}{3}(h_0^2-m^2)^{3/2}+\frac{2h_0^2-m^2}{h_0(h_0^2-m^2)}\frac{\Delta_0^2}{\lambda}\bigg).\ee 
Substituting in the the relations between
  $g_0,h_0,\Delta_0$ and $z$ from the near horizon
  EOM, it is easy to show that % for both the electron star and BCS star
  %cases 
$c_0=2z^2$ and does not depend on the BCS coupling
  $\lambda$. Hence index of the Hankel function is just $3/2$.

Near the AdS$_4$ boundary, on the other hand, we have
\be
\delta a_x=\delta a_x^{(0)}+\frac{\delta a_x^{(1)}}{r}+\dots,
\ee
The conductivity for the dual field theory is extracted from these values as
\be
\label{cond}
\sigma=-\frac{i}{\omega}\frac{\delta a_x^{(1)}}{\delta a_x^{(0)}}. 
\ee

For low frequencies the near AdS$_4$ boundary coefficients
  can be related to the near horizon behavior through
% Following \cite{Hartnoll:2010gu} we consider
the conserved quantity  \cite{Hartnoll:2010gu} %for (\ref{eomax}) 
\be
\mathcal{F}=i\sqrt{\frac{f}{g}}\bigg(\delta a_x^*\partial_r\delta a_x-
\delta a_x\partial_r\delta a_x^*\bigg).
\ee
% It has the following properties.
Near the AdS$_4$ boundary $\cF$ equals $\mathcal{F}=-2\omega|\delta
a_x^{(0)}|^2 \text{Re} \sigma$, whereas the near horizon solution (\ref{nhdeltaax}) 
gives $\mathcal{F}\sim\omega% \commentr{|\delta a_x^{(\text{in})}|^2}
$. Thus
$\text{Re}\sig \sim |\delta
a_x^{(0)}|^{-2}$.  %\commentr{(|\delta a_x^{(\text{in})}|/|\delta
%a_x^{(0)}|)^{2}}$. % For low frequency region, 
Using the matching method it is then easy to see $|\delta
a_x^{(0)}|\propto \omega^{-1}$ %|\delta a_x^{(\text{in})}|$ 
\cite{Hartnoll:2010gu}.
Thus the low frequency behavior of the conductivity is 
\be
\text{Re}~\sigma\propto\delta(\omega)+\omega^{2}.\ee
Notice that the delta function has to be there due to the
  translation invariance of the BCS star background. In more detail, this arises 
  from the pole in the conductivity  when $\omega\to 0$, as can be verified by evaluating (\ref{cond}) explicitly.

The small frequency behavior  of the conductivity % are
is in fact independent of the BCS coupling $\lambda$, since the computation
is in this regard tracks not different  from the computation for the electron star with
$\lambda=0$. This hard gap is also missing in the holographic
superconductor \cite{Gubser:2009cg,Horowitz:2009ij}. 
This is understood as a remnant effect of the near-horizon Lifshitz
geometry. Since the geometry persists all the way to $r\rar 0$, in the
dual field theory there are a ``large $N$'' amount of degrees of
freedom surviving in the IR. These coexists with the phase mode of 
the superconductor,  causing the remnant finite conductivity
in the region where which would be fully gapped in a conventional superconductor.

%\comment{I do not understand either of these sentences FIX OR DROP: Here the
%  missing of this hard gap is related to the fact that the gap in the
%  spectral function is not a hard gap.
%\comment{Is this sentence correct?} 
%The spectral function is exactly zero at zero frequency \cite{Liu:2012tr} so there should be a $\delta$ function %at zero frequency. At nonzero frequency, there is a small nonvanishing spectral weight.}

% We cannot see a hard gap in the conductivity, but it is expected that a hard gap will arise after taking into consideration the quantum corrections. 

%\comment{why no gap?}

%%%%%%%%%%%%%%%%%%%%%%%%%%%%%%%%
\section{Scaling limits with a dynamical scalar}
\label{seccom}
%%%%%%%%%%%%%%%%%%%%%%%%%%%%%%%%

In our BCS star model, the matter fields are not visible at the
boundary, though in the last section we % already
showed that there are still effects on the boundary theory. Here % in this section
we study a more generalized model % with the inclusion of
which includes dynamics for the scalar field $\Delta$. 
Technically this will allow $\Delta$ to extend all the way to boundary.
% a dynamical scalar field and show that
Physically, from the pure BCS perspective, this may seem strange. Indeed the most
natural way to interpret the dual field theory this model describes, is as a system with charged
fermions and an additional independent charged scalar operator with
charge $q_{b}=2q_f$. From the gravity perspective, however,
it is a very natural description that arises in many top-down
models.
% A dynamical scalar field also allows for a comparison of the BCS star with standard
% holographic superconductors \cite{Hartnoll:2008vx,Faulkner:2009am}; each arises as a special limit of this more generalized model.

%%%%%%%%%%%%%%%%%%%%%%%%%%%%%%%%
\subsection{Lagrangian with a dynamical scalar}
%%%%%%%%%%%%%%%%%%%%%%%%%%%%%%%%

In \cite{Liu:2013yaa} we considered models with both
  dynamical scalars and fermions. There we found that the holographic description of strongly
coupled systems with both bosons and fermions with incommensurate
charges $q_b\neq 2n q_f,~n \in \mathbb{N}$, has electron star solutions which
can coexist with scalar hair (see also \cite{Nitti:2013xaa}). This corresponds to a superconducting
state with multiple Fermi surfaces. In that case the incommensurate charge prevents
a relation between the fermions and bosons in the gravitational bulk, 
and hence in the boundary.
% It is natural to specially focus on the case that the bosons can be relevant to
% the condensation of the fermions. In this case $q_b=2 q_f$ and we can
% introduce a bulk Yukawa/BCS interaction between fermions and
% bosons. 

A commensurate scalar charge $q_b=2 q_f$ allows a
  Yukawa/BCS interaction between fermions and bosons in the bulk. This
is what we studied so far, but without explicit dynamics for the
scalar field. It only arose as an auxiliary field.
For a dynamical scalar, on the other hand, it is natural to surmise that the energetics
of the bosons can be relevant to
 the condensation of the fermions.
 % direct 
The more generalized system we % are
therefore consider % in this section
is  % the following Lagrangian:
\bea \mathcal{L}&=&\frac{1}{2\kappa^2}\bigg(R+\frac{6}{L^2}\bigg)-\frac{1}{4e^2}F_{\mu\nu}F^{\mu\nu}
-|(\partial_\mu-2iqA_\mu)\phi|^2-m_\phi^2|\phi|^2\nonumber\\ &&
-i\bar\Psi(\Gamma^\mu  \mathcal{D}_\mu-m_\Psi)\Psi
+\eta_5^*\phi^*\bar\Psi_c\Gamma^5\Psi-\eta_5\phi\bar\Psi \Gamma^5\Psi_c.
\eea

This model has been considered before in
  \cite{Faulkner:2009am} from a perspective where the fermions are probes,
whereas $\eta_5=0$ this is a special case of the bose-fermi
  competition models studied in \cite{Liu:2013yaa,Nitti:2013xaa} with
$q_b=2q_f$. %\commentr{
Its connection to the BCS Lagrangian studied
  here is made clear after the field redefinition
\be
\phi= \frac{1}{m_{\phi}\sqrt{2\lambda}}\Delta, ~~~ \eta_5 =
m_{\phi}\sqrt{\frac{\lambda}{2}}. %m_\phi=2\lambda_0 m_\Delta.
\ee
%\comment{I am happy to revert back to the old notation, if you can
%  explain why it is better.}
Then the Lagrangian becomes:
\bea \mathcal{L}&=&\frac{1}{2\kappa^2}\bigg(R+\frac{6}{L^2}\bigg)-\frac{1}{4e^2}F_{\mu\nu}F^{\mu\nu}
-\frac{1}{2\lambda m_{\phi}^2}|(\partial_\mu-2iqA_\mu)\Delta|^2-\frac{1}{2\lambda}|\Delta|^2\nonumber\\ &&
-i\bar\Psi(\Gamma^\mu  \mathcal{D}_\mu-m_f)\Psi
+\frac{1}{2}\Delta^*\bar\Psi_c\Gamma^5\Psi-\frac{1}{2}\Delta\bar\Psi \Gamma^5\Psi_c. 
\eea
In the formal limit $m_{\phi}^2 \rar \infty$ we recover the
Einstein-Maxwell-BCS Lagrangian. We will now make this limit more
precise.%}
%\comment{OLD $4\lambda_0^2=2\lambda m_{\phi}^2$; OLD $m_{\Delta}^2 = 1/2\lambda$}
 
The equations of motion for this system are
\bea
R_{\mu\nu}-\frac{1}{2}g_{\mu\nu}R-\frac{3}{L^2}g_{\mu\nu}-\kappa^2\bigg[T_{\mu\nu}^{\text{gauge}}+T_{\mu\nu}^{\text{BCS}}\bigg]&=&\kap^2\bigg[T_{\mu\nu}^{\text{kin.boson}}\bigg];\nonumber\\
{\nabla_\mu}F^{\mu\nu}+e^2J^{\nu}_{\text{BCS}}&=&% \frac{iqe^2}{2\lambda_0^2}
\frac{iqe^2}{\lambda
  m_{\phi}^2}\bigg[\Delta^*\big(\partial^\nu-2iqA^\nu\big)\Delta-\Delta\big(\partial^\nu+2iqA^\nu\big)\Delta^*\bigg];\nonumber\\
\Delta-\lambda\bar\Psi_c\Gamma^5\Psi&=&\frac{1}{m_{\phi}^2}\big(\nabla^\mu-2iqA^\mu\big)\big(\nabla_\mu-2iqA_\mu\big)\Delta; \label{eom222} \\
i\big(\Gamma^{\mu}\mathcal{D}_{\mu}-m_f\big)\Psi-\Delta^\dagger 
\Psi_c\Gamma^5\Psi&=&0,\nonumber
\eea
where $T_{\mu\nu}^{\text{gauge}}$, $T_{\mu\nu}^{\text{BCS}}$ and $J^\mu_{\text{BCS}}$ are as
before in Eqns \eqref{emcase1} and \eqref{totalfluid}, and 
\bea
% T_{\mu\nu}^{\text{gauge}}&=&\frac{1}{e^2}\bigg(F_{\mu\rho}F_{\nu}^{~\rho}-\frac{1}{4}F^2g_{\mu\nu}\bigg);\nonumber\\
% T_{\mu\nu}^{\text{BCS}}&=&
% \frac{1}{2}\langle-i\bar\Psi\Gamma_{(\mu}\mathcal{D}_{\nu)}\Psi+i\bar\Psi \overlef\label{eom222}tarrow{\mathcal{D}}_{(\mu}\Gamma_{\nu)}\Psi\rangle\nonumber\\
% &&~~
% -g_{\mu\nu}\bigg(\langle% \bigg[
% -i\bar\Psi(\Gamma^\alpha\mathcal{D}_{\alpha}-m_{\Psi})\Psi+\frac{1}{2}\Delta^*\bar\Psi_c\Gamma^5\Psi+\Delta\bar\Psi \Gamma^5\Psi_c % \bigg]
% \rangle
% +m_\Delta^2|\Delta|^2\bigg)\nonumber\\
%&=&
%(\rho+p)u_\mu u_\nu+pg_{\mu\nu},\nonumber\\
%\bar\Psi \overleftarrow{\mathcal{D}}_\mu&=&\partial_\mu\bar{\Psi} +\frac{1}{4}\omega_{ab \mu}\bar{\Psi} \Gamma^{ab}+iqA_\mu\bar{\Psi} ;\nonumber\\
% T_{\mu\nu}^{\text{kin.boson}}&=&\frac{1}{4\lambda_0^2}(\partial_\mu+2iqA_\mu)\Delta^*(\partial_\nu-2iqA_\nu)\Delta+\frac{1}{4\lambda_0^2}(\partial_\mu-2iqA_\mu)\Delta(\partial_\nu+2iqA_\nu)\Delta^*\nonumber\\&&~~-g_{\mu\nu}(\frac{1}{4\lambda_0^2}|(\partial_\alpha-2iqA_\alpha)\Delta|^2);\nonumber\\
T_{\mu\nu}^{\text{kin.boson}}&=&\frac{1}{\lambda m_{\phi}^2}\bigg(\big(\partial_{(\mu}+2iqA_{(\mu}\big)\Delta^*\big(\partial_{\nu)}-2iqA_{\nu)}\big)\Delta % +\frac{1}{4\lambda_0^2}(\partial_\mu-2iqA_\mu)\Delta(\partial_\nu+2iqA_\nu)\Delta^*\nonumber\\&&~~
  -\frac{1}{2}g_{\mu\nu}|\big(\partial_\alpha-2iqA_\alpha\big)\Delta|^2\bigg),%\nonumber\\
% J^\mu_{\text{fermion}}&=&q n u^{\mu}=-q\langle\bar{\Psi}\Gamma^\mu\Psi\rangle,
\eea
with $A_{(\mu}B_{\nu)}=\frac{1}{2}(A_\mu B_\nu+A_\nu B_{\mu})$. The
terms on the right hand side of \eqref{eom222} are new compared to the pure BCS system
considered before. %and $u^2=1$, $u_t=e_{t\underline{t}}$.
The decoupling limit needs more in depth inquiry, because we
  first need to impose a well-defined semi-classical limit for the
  many body fermion system. Making the fluid approximation $T_{\mu\nu}^{\text{BCS}}=
(\rho+p)u_\mu u_\nu+pg_{\mu\nu}$ as in \eqref{totalfluid}, this is obtained in terms of the
dimensionless variables found earlier
\be\label{scaling2}
(\rho,p)=\frac{1}{\kappa^2}(\hat{\rho},\hat{p}),~~~n=\frac{1}{e\kappa}\hat{n},~~~
(A_\mu,\mu_{l},m_f,\Delta,\omega_D)=\frac{e}{\kappa}(\hat{A}_{\mu},\hat{\mu},\hat{m}_f,\hat{\Delta},\hat{\omega}_D),~~~%m_\Delta=\frac{1}{\sqrt{2}e}\frac{1}{\hat{\lambda}}
\lambda= \frac{e^2}{\beta}\hat{\lambda}
\ee
%\comment{We need to explain why $\Delta$ does not include $\beta$ in the redefinition (2.41)}. 
where the hatted quantities are of order zero in $\kappa$ and $e$
with $\beta = e^4/\pi^2\kap^2$ fixed,
and for simplicity we have set $L=1$ and $q=1$ as $q$ only appears in the
combination $q e$. In terms of the rescaled variables the bosonic EOM become
\bea
R_{\mu\nu}-\frac{1}{2}g_{\mu\nu}R-3 g_{\mu\nu}-\bigg[\hat{T}_{\mu\nu}^{\text{gauge}}+\hat{T}_{\mu\nu}^{\text{BCS}}\bigg]&=&\frac{\beta}{\hat{\lambda} m_{\phi}^2}\bigg[ (\partial_{(\mu}+2i\frac{q_{\text{eff}}}{\sqrt{\kap}}\hat{A}_{(\mu})\hat{\Delta}^*(\partial_{\nu)}-2i\frac{q_{\text{eff}}}{\sqrt{\kap}}\hat{A}_{\nu)})\hat{\Delta}\nonumber\\
&&~~~-\frac{1}{2}
g_{\mu\nu}|\big(\partial_\alpha-2i\frac{q_{\text{eff}}}{\sqrt{\kap}}\hat{A}_\alpha\big)\hat{\Delta}|^2
\bigg];\nonumber\\
{\nabla_\mu}\hat{F}^{\mu\nu}+\hat{J}^{\nu}_{\text{BCS}}&=&
\frac{e}{\kappa}\frac{i\beta}{\hat{\lambda}
  m_{\phi}^2}\bigg[\hat{\Delta}^*(\partial^\nu-2i\frac{q_{\text{eff}}}{\sqrt{\kap}}\hat{A}^\nu)\hat{\Delta}-\hat{\Delta}(\partial^\nu+2i\frac{q_{\text{eff}}}{\sqrt{\kap}}\hat{A}^\nu)\hat{\Delta}^*\bigg];\nonumber\\
\hat{\Delta}-\hat{\lambda}\langle\bar\Psi_c\Gamma^5\Psi\rangle&=&\frac{1}{m_{\phi}^2}(\nabla^\mu-2i\frac{q_{\text{eff}}}{\sqrt{\kap}}\hat{A}^\mu)(\nabla_\mu-2i\frac{q_{\text{eff}}}{\sqrt{\kap}}\hat{A}_\mu)\hat{\Delta}, 
\eea
where $q_{\text{eff}}= \sqrt{\pi}\beta^{1/4}$.

We see that there is no clean 
% To solve this system in the
classical gravity limit $\kappa \to 0$, where the rescaled fields can
stay fixed and
the energy momentum contribution to the gravity is still of order $\mathcal{O}(1)$. 
This is precisely due to the fact that the bosonic charge is fixed in
units of the fermion charge. For incommensurate charges, i.e. if
$q_{\text{eff}}$ were a free parameter, one can scale this charge to
absorb the explicit dependence on the gravitational coupling $\kappa$; see \cite{Liu:2013yaa}.
The fact that the fluid limit is incompatible with a scaling
limit in the microscopic Lagrangian was already noted in
\cite{Cubrovic:2011xm}.

In our case, where $q_b$ is not free, but fixed to equal $q_b=2q_f$,
there are three possible classical limits. They depend on the scaling
choice for the mass $m_{\phi}$. One has:

\begin{itemize}

\item $m_{\phi}^2=\kappa^{-1-\delta}\hat{m}_{\phi}^2$ where $\delta>0$:
  This is the limit where the kinetics of the scalar completely
  decouples and one recovers the system studied in the previous sections. 

\item $m_{\phi}^2=\kap^{-1}\hat{m}_{\phi}^2$. This is the natural
  limit in which the hatted parameter $m_{\phi}^2$ is a truly dimensionless
  parameter. In this limit the strict kinetics of the scalar field are unimportant, but the coupling to the gauge field and to the fermionic field remain. This is exactly the case we will study in this section.

\item  $m_{\phi}^2=\kappa^{-1-\delta}\hat{m}_{\phi}^2$ where
  $\delta<0$. This is not a well defined classical limit which means the scaling (\ref{scaling2}) could be modified resulting in that not all fermionic terms could be kept. Applying it nevertheless means that 
the kinetics of the scalar field can be kept and dominate but its derivative decouples from the
Maxwell connection. In essence $q_{\text{eff}}$ must be set to
zero. We leave this case for future study.
\end{itemize}

%%%%%%%%%%%%%%%%%%%%%%%%%%%%%
\subsection{Charged non-dynamical scalar scaling limit}
%%%%%%%%%%%%%%%%%%%%%%%%%%%%%

We now focus on the second case where $m_{\phi}^2
  =\kap^{-1}\hat{m}_{\phi}^2$ and take the limit $\kap \rar 0$ with
  all hatted quantities fixed. The ansatz for the background we take is the same as
(\ref{bggauan}). % Similar to the previous case we define 
We now define a new combined fluid
$$T_{\mu\nu}^{\text{BCS}}+T_{\mu\nu}^{\text{kin.boson}}=(\rho_{\text{com}}+p_{\text{com}}) u_\mu u_\nu+p_{\text{com}} g_{\mu\nu}$$
and $$J_{\mu}^{\text{fermion}}+J_{\mu}^{\text{boson}}=n_{\text{com}}
u_\mu$$
where the rescaled fluid quantities
%\be 
%\hat{\rho}_{\text{com}}=\hat{\rho}+\frac{1}{\hat{m}_{\phi}^2}\frac{2\pi\beta^{3/2} h^2}{
%  f}\frac{\hat{\Delta}^2}{\hat{\lambda}},~~~
%\hat{p}_{\text{com}}=\hat{p}+\frac{1}{\hat{m}_{\phi}^2}\frac{2\pi\beta^{3/2}
%  h^2}{f}\frac{\hat{\Delta}^2}{\hat{\lambda}},~~~\hat{n}_{\text{com}}=\hat{n}+\frac{1}{\hat{m}_{\phi}^2}\frac{4\pi\beta^{3/2}
%  h}{\sqrt{f}}\frac{\hat{\Delta}^2}{\hat{\lambda}}
%\ee
\be 
\hat{\rho}_{\text{com}}=\hat{\rho}+\frac{s\beta h^2}{
  f}\frac{\hat{\Delta}^2}{\hat{\lambda}},~~~
\hat{p}_{\text{com}}=\hat{p}+\frac{s\beta
  h^2}{f}\frac{\hat{\Delta}^2}{\hat{\lambda}},~~~\hat{n}_{\text{com}}=\hat{n}+\frac{2s\beta
  h}{\sqrt{f}}\frac{\hat{\Delta}^2}{\hat{\lambda}}
\ee
% \be 
% \rho_{\text{com}}=\hat{\rho}+\frac{\beta h^2}{\hat{\lambda}^2
%   f}\hat{\Delta}^2,~~~p_{\text{com}}=\hat{p}+\frac{\beta
%   h^2}{\hat{\lambda}^2
%   f}\hat{\Delta}^2,~~~n_{\text{com}}=\hat{n}+\frac{2\beta
%   h}{\hat{\lambda}^2 \sqrt{f}}\hat{\Delta}^2
% \ee
% with $(\hat{\rho},\hat{p},\hat{n})$ the rescaled 
are the 
BCS fluid quantities in (\ref{totalfluid}) and we have 
  introduced the parameter $s$ related to the scalar mass % is introduced
  for convenience
\be
s\equiv \frac{2\pi \sqrt{\beta}}{\hat{m}_\phi^2}. 
\ee
Obviously, when $s\to 0$, i.e. $\hat{m}_\phi^2\to\infty$, our system
reduces to the BCS star system discussed in the previous section.  
For finite $s$ the equations of motion for the system in terms of the combined fluid 
are the same as the previous
case, (\ref{eomsystem}-\ref{eomec}) % by replacing $(\hat{\rho},\hat{p},\hat{n})$ to $(\rho_{\text{com}},p_{\text{com}},n_{\text{com}})$ with 
% \be \rho_{\text{com}}=\hat{\rho}+\frac{\beta h^2}{\hat{\lambda}^2 f}\hat{\Delta}^2,~~~p_{\text{com}}=\hat{p}+\frac{\beta h^2}{\hat{\lambda}^2 f}\hat{\Delta}^2,~~~n_{\text{com}}=\hat{n}+\frac{2\beta h}{\hat{\lambda}^2 \sqrt{f}}\hat{\Delta}^2\ee
% where $(\hat{\rho},\hat{p},\hat{n})$ are rescaled quantities in (\ref{totalfluid}, \ref{rescalesecII}).
%The extra terms here are from the interaction between the scalar field and the Maxwell field.
with the exception of 
the equation of motion for the scalar field. It gives 
\be
-\frac{2s h^2}{f\hat{\lambda}}+\frac{1}{\hat{\lambda}}=\pi^2\int_{-\hat{\omega}_D}^{\hat{\omega}_D}d\xi\hat{\nu}(\xi)\frac{1}{\sqrt{\xi^2+\hat{\Delta}^2}}.
\ee 
In the same limit as before ${\hat{\Delta}}\ll\hat{\omega}_D\ll (\hat{\mu},\hat{\mu}-\hat{m}_f)$,
this modified gap equation can be solved as 
\be\label{newdelta}
\hat{\Delta}=2\hat{\omega}_D e^{-(1-\frac{2sh^2}{f})/(\hat{\lambda}\hat{\mu}\sqrt{\hat{\mu}^2-\hat{m}_f^2})}.
\ee 
From (\ref{newdelta}) it is easy to see when $s$ is large, $1-2s h^2/f$ would be negative and the approximation ${\hat{\Delta}}\ll\hat{\omega}_D$
would break down. Thus % our formula
the perturbative approach we follow here only applies for small $s$.

Let us explain in more detail the way this system works in this limit, where
especially the role of the scalar field equation is interesting.
In this limit  all kinetics decouple: the scalar field has become an
auxiliary field again. However, we can see that $\Delta$ is no longer
the Cooper pair condensate. Nevertheless,  this gap is still associated 
with  a local superconducting state in  the  bulk as can be seen from the
Dirac equation. What is happening is that the charged gap field is now
also sensitive to the background gauge connection. Note that it does do so in a
way that gauge symmetry is broken. The gap field has the status of a
Stueckelberg field. The limit is therefore a Stueckelberg limit where
strict decoupling does not happen. Only at low energies, much below
$m_{\phi}$,  this is a reliable approximation to the system.

\begin{figure}[t!]
\begin{center}
%\begin{tabular}{cc}
%\includegraphics[width=0.45\textwidth]{fluid-new.pdf}
%\includegraphics[width=0.45\textwidth]{delta-new.pdf}
%\end{tabular}
\begin{tabular}{cc}
\includegraphics[width=0.43\textwidth]{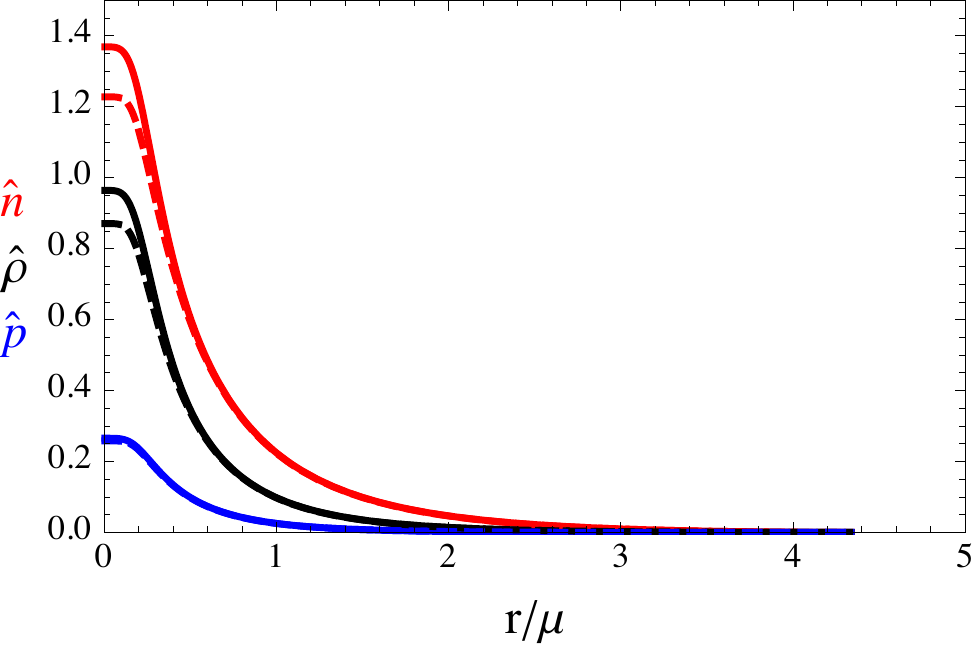}
\includegraphics[width=0.45\textwidth]{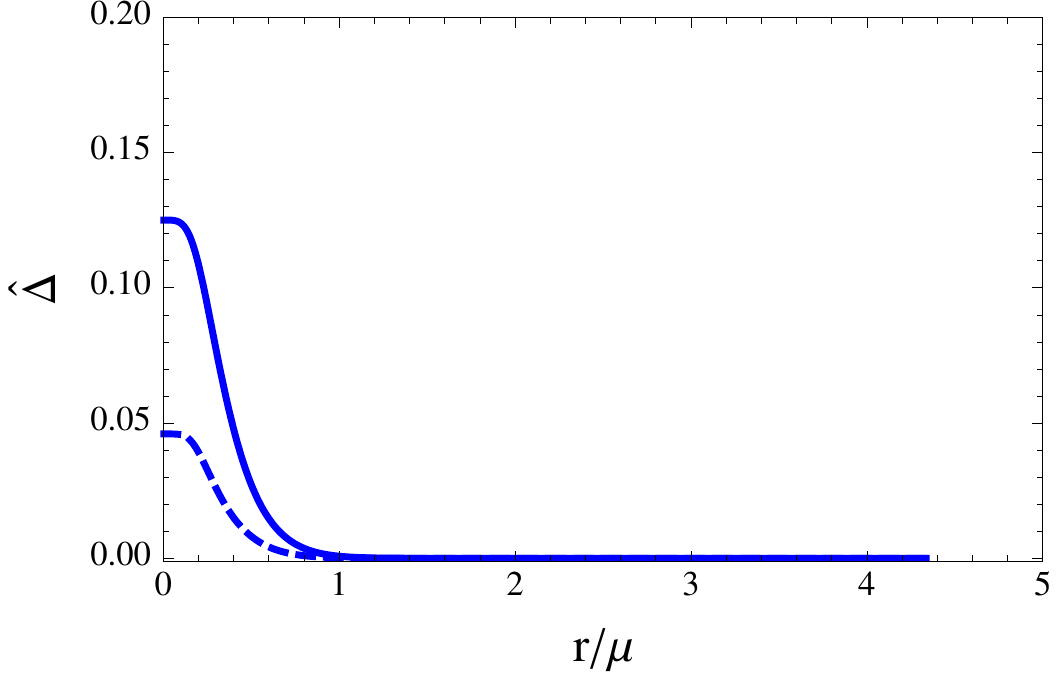}
\end{tabular}
\caption{\small % Up: The profile for the fluid in the star in the
% BCS-Stueckelberg limit as a function of the radial
%   coordinate with $\hat{m}_f=0.2, c=1/3, \beta=5$,
%   $\hat{\lambda}\hat{\mu}\sqrt{\mu^2-\hat{m}^2}=$ 0.393 and $s=0.25$. 
%  Left:
%   from top to bottom, the solid lines are $\hat{n}_\text{com}$,$\hat{\rho}_\text{com}$,$\hat{p}_\text{com}$ while the dashed lines are $\hat{n}_\text{tot}$,$\hat{\rho}_\text{tot}$,$\hat{p}_\text{tot}$; Right:
%   $\hat{\Delta}$.
% \comment{Can we plot this in comparison to the $m_{\phi}=\infty$
%   case?---is it meaningful? which one should we keep, up or down? }
  % Down:
  The profile for the fluid in the star in the
BCS-Stueckelberg limit as a function of the radial
  coordinate with $\hat{m}_f=0.2, c=1/3, \beta=5$,
  $\hat{\lambda}\hat{\mu}\sqrt{\mu^2-\hat{m}_f^2}=$ 0.393. 
 Left:
  from top to bottom, the solid lines are
  $\hat{n}_\text{com}$,~$\hat{\rho}_\text{com}$,~$\hat{p}_\text{com}$
  with $s=0.25$ and the star edge $r_s/\mu\simeq 4.338$. For comparison, we also give the profiles of the pure
  BCS star
  $\hat{n}_\text{tot}$,~$\hat{\rho}_\text{tot}$,~$\hat{p}_\text{tot}$
  with $s=0$ and the star edge $r_s/\mu\simeq 4.328$; Right: The value of the gap
  $\hat{\Delta}$ for $s=0.25$ (solid) and $s=0$ (dashed) for the same
  numerical parameters.
Both the gap and the charge density are enhanced compared to the pure
BCS star.
}
\label{fluidnew}
\end{center}
%\end{figure}
%%%%%%%%%%%%%%%%%%%%%%%%%%%%%%%%%%
%%%%%%%%%%%%%%%%%%%%%%%%%%%%%%%%%%
%\begin{figure}[h]
\begin{center}
\begin{tabular}{cc}
\includegraphics[width=0.5\textwidth]{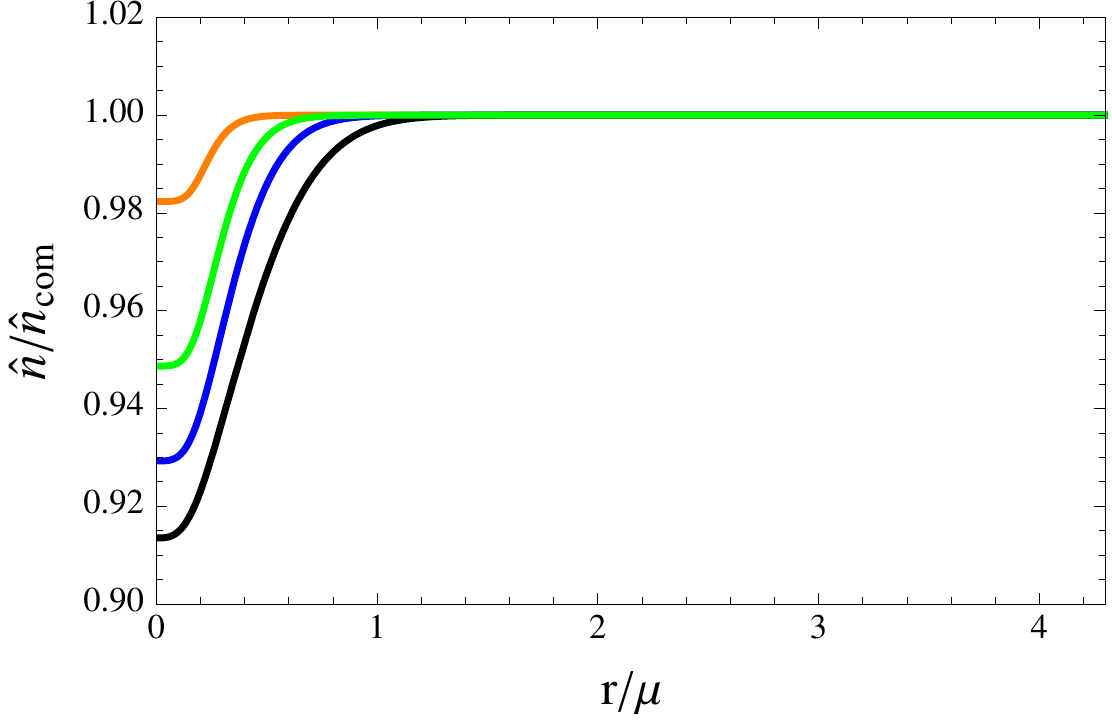}
\end{tabular}
\caption{\small The ratio of the pure BCS charge density $\hat{n}$ to
  the combined BCS-Stueckelberg charge density $\hat{n}_{\text{com}}$ as a function the radial coordinate for different coupling constant $\lambda$ for $\hat{m}_f=0.2, c=1/3, \beta=5$ and  
$s=0.25$. In the figure, $\hat{\lambda}\hat{\mu}\sqrt{\mu^2-\hat{m}_f^2}=$ 0.245 (Orange), 0.393 (Green), 0.534 (Blue),  0.810 (Black).}
\label{nfnnew}
\end{center}
\end{figure}

\begin{figure}[t!]
\begin{center}
\begin{tabular}{cc}
\includegraphics[width=0.45\textwidth]{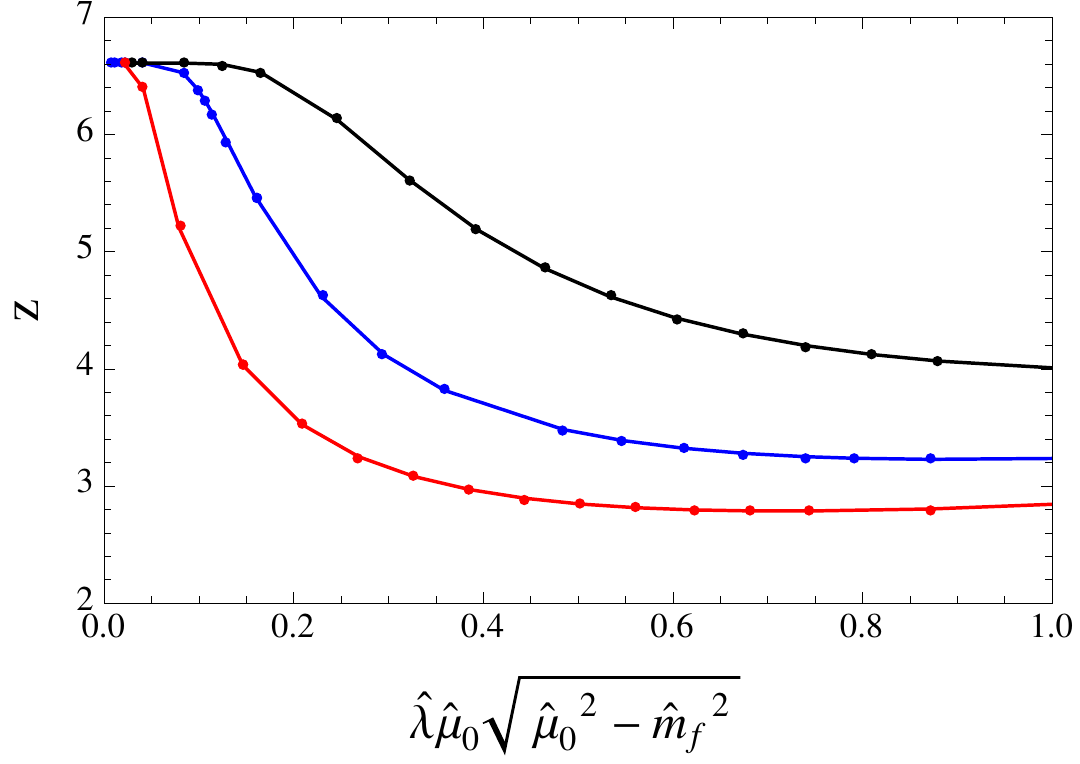}
\raisebox{-.05in}{\includegraphics[width=0.5\textwidth]{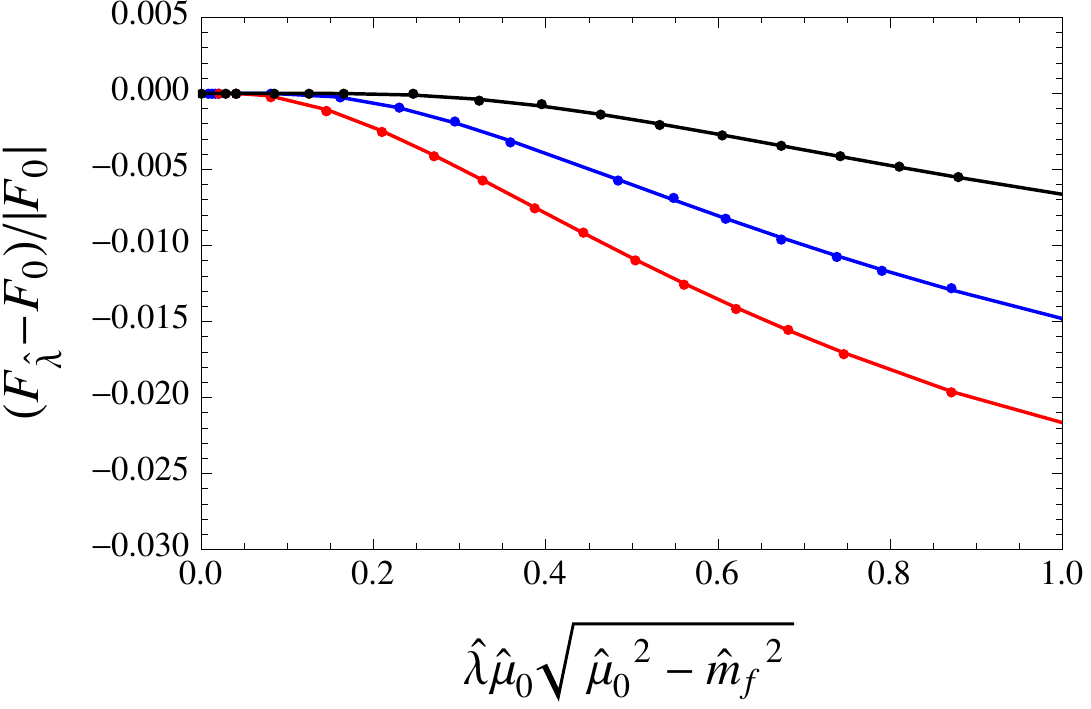}}
\end{tabular}
\caption{\small The near horizon Lifshitz scaling $z$ ({left} plot) and free energy ({right} plot) for the star in this new scaling limit as a function of $\lambda$ for $\hat{m}_f=0.2,  c=1/3,\beta=5$ with different  
$s=0.25$ (black), $0.4$ (blue), $0.5$ (red). 
The free energy decreases as $\hat{\lambda}$ increases in the region
that perturbation theory applies. The larger the Stueckelberg term, the
more thermodynamically favored the solution is (and the smaller the IR
dynamical critical exponent). This is in accordance
with the corresponding increase in
the gap.}
\label{fznew}
\end{center}
\end{figure}

For the solution of this BCS-Stueckelberg system, we
  proceed as before. 
The near horizon geometry is still Lifshitz. One can add an irrelevant perturbation for the geometry to flow to an AdS solution. The behavior 
of the fluid and condensate is plotted in
Fig. \ref{fluidnew}. The significant difference compared to
  the pure BCS star is the enhancement in the charge density
  (Fig. \ref{nfnnew}). In particular we see 
  the BCS-Stueckelberg star is more susceptible to form a
  superconducting core. This can be directly understood from the
  reduced suppression of the gap. The stronger predilection towards
  pairing should also be reflected in the thermodynamic
  properties. Indeed the BCS-Stueckelberg star in this
limit is more stable (Fig. \ref{fznew}).

%%%%%%%%%%%%%%%%%%%%%%%%%%%%%%%%%%
%%%%%%%%%%%%%%%%%%%%%%%%%%%%%%%%%%

%%%%%%%%%%%%%%%%%%%%%%%%%%%%%%%%%%
%%%%%%%%%%%%%%%%%%%%%%%%%%%%%%%%%%

% The clearest distinction between the pure BCS and the
%   BCS-Stueckelberg star is that now the charges are 
% carried by both fermions and bosons.
% When $\hat{\lambda}$ increases one can see from fig. \ref{nfnnew} that $\hat{n}_f/\hat{n}_t$ decreases which means more fermions form pairs. As to the first scaling case, because all the charges are 
%carried by fermions and $\hat{n}_f/\hat{n}_t=1$, it is not quite
%natural to conclude more bcs  pairing are formed when we tune
%$\lambda$. 

\begin{figure}[t!]
\begin{center}
\begin{tabular}{cc}
\includegraphics[width=0.50\textwidth]{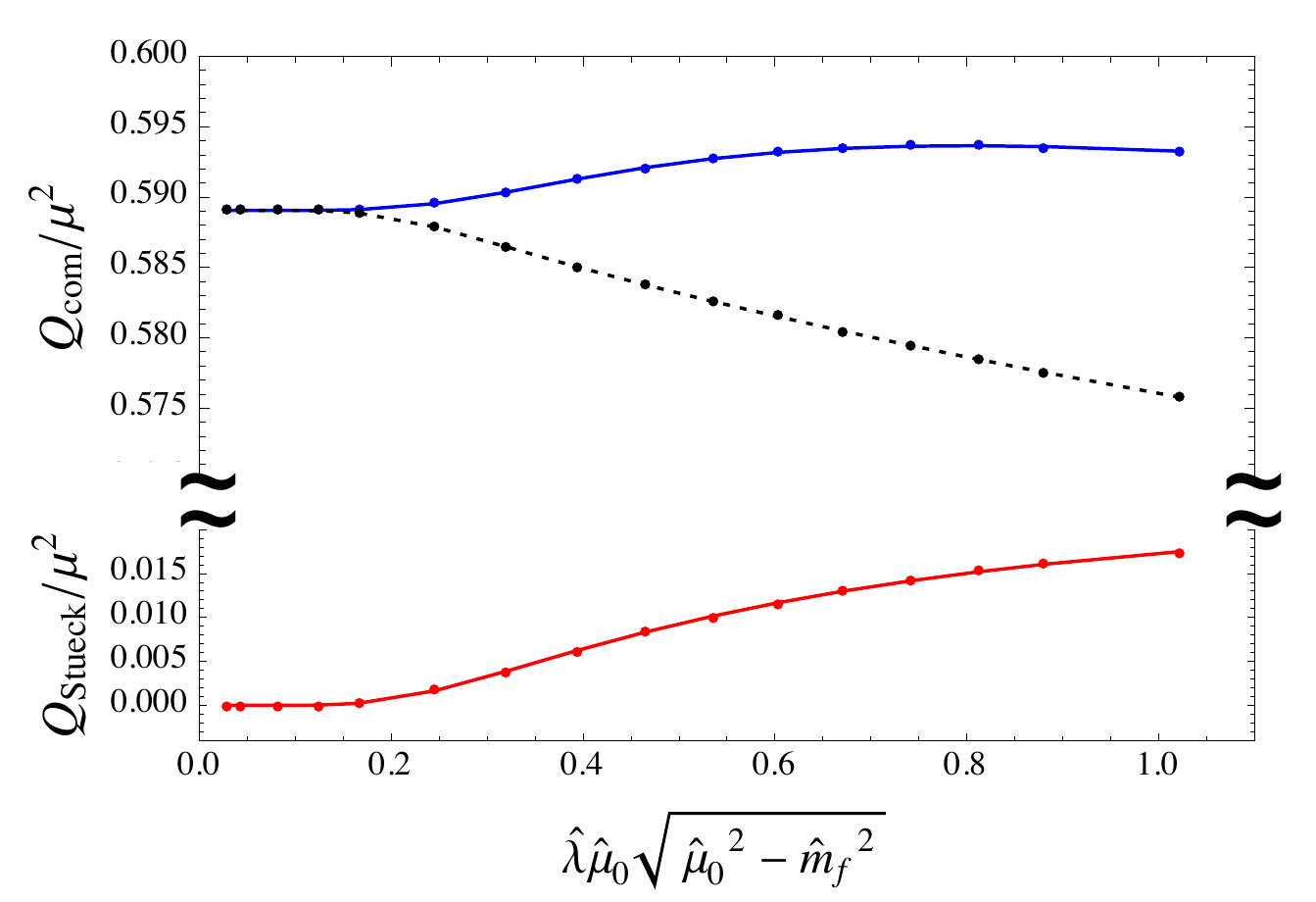}
\raisebox{-.05in}{\includegraphics[width=0.49\textwidth]{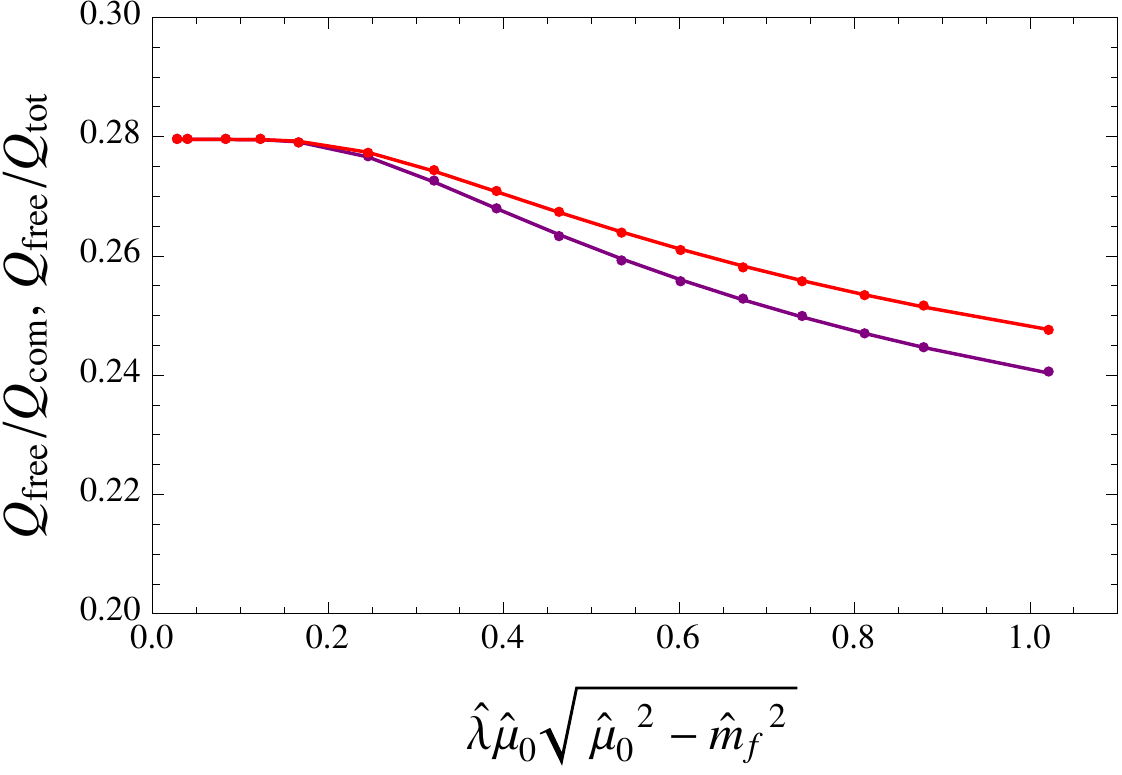}}
%\raisebox{-.14in}{\includegraphics[width=0.49\textwidth]{ratio-charge-ks-v2.pdf}}
\end{tabular}
\caption{\small Left: The absolute charge densities
    $Q_{\text{com}}/\mu^2$ (blue), $Q_{\text{tot}}/\mu^2$ (black dashed) and
    $Q_{\text{Stueck}}/\mu^2$ (red) in the BCS-Stueckelberg
    star as a function of the coupling  $\hat{\lambda}\hat{\mu}_0\sqrt{\hat{\mu}_0^2-\hat{m}_f^2}$. Right: The relative charge densities of
    $Q_{\text{free}}/Q_{\text{tot}}$ (red) and
    $Q_{\text{free}}/Q_{\text{com}}$ (purple) as a function of the coupling $\hat{\lambda}\hat{\mu}_0\sqrt{\hat{\mu}_0^2-\hat{m}_f^2}$ for $\hat{m}_f=0.2, c=1/3, \beta=5$ and $s=0.25.$}
\label{ratio-secIV}
\end{center}
\end{figure}

The total charge distributions are also reflecting this extra stability. In the BCS-Stueckelberg star we can distinguish a third
  component contributing to the charge density: next to the free- and paired fermions there is also the contribution from 
  the Stueckelberg field. 
% There are therefore three kinds of charge carriers:
% We can see that as the coupling becomes larger, there will be more
% bosons. More bosons will also cause more fermions to pair. To show
% this in a more clear way, we integrate the bulk charge density along
% the radial direction which gives the corresponding boundary charge
% density. We have \be
Define a new combined charge density by
\be
Q_{\text{com}}=\int_0^{r_s} dr r^2\sqrt{g_{rr}} n_{\text{com}}
\ee
in addition to the densities $Q_{\text{free}}$ and $Q_{\text{total}}$
as  given by (\ref{QQinSecII}). We  can then define the Stueckelberg charge density as
$Q_{\text{Stueck}}=Q_{\text{com}}-Q_{\text{total}}$. 
%and here $Q_{\text{free}}$ is the same as (\ref{QQinSecII}).
The left plot of Fig. \ref{ratio-secIV} demonstrates that  this extra
Stueckelberg contribution gives rise to an increase of the charge density 
upon increasing the BCS coupling, as expected intuitively. 
Whereas the pure BCS contribution $Q_{\text{total}}$  decreases with increasing coupling
as before, the extra Stueckelberg
contribution suffices  to compensate for the depletion of the
free fermionic density, as illustrated in the right plot of Fig. \ref{ratio-secIV}. 

Finally, we checked by explicit calculation along the lines of the previous section
that the gap in the dual Fermion spectral function continues to be set by  $\Delta$
also in this  BCS-Stueckelberg limit. The novelty is just  that the Stueckelberg field is 
enhancing this gap. 

%%%%%%%%%%%%%%%%%%%%%%%%%%%%%%%%%%
%%%%%%%%%%%%%%%%%%%%%%%%%%%%%%%%%%

%%%%%%%%%%%%%%%%%%%%%%%%%%%%%%%%%%%%%%%
%%%%%%%%%%%%%%%%%%%%%%%%%%%%%%%%%%%%%%%
\section{Conclusion and discussion}
\label{sec5}
%%%%%%%%%%%%%%%%%%%%%%%%%%%%%%%%%%%%%%%
%%%%%%%%%%%%%%%%%%%%%%%%%%%%%%%%%%%%%%%

In this paper we have made a step towards understanding fermion driven
pairing in strongly coupled systems with holographic duals. In
particular we considered the introduction of a BCS interaction
for the fields {\em dual} to the fermionic operators in strongly coupled
theory, i.e. we complemented the AdS-Einstein-Maxwell-Dirac action
with a standard BCS interaction. This implicitly assumes that
at low energies these fermionic operators control the physics and that
the pairing is driven by a force other than the one that controls the
strong correlations, even
though this might 
be unnatural from more microscopic arguments or top-down AdS/CFT
constructions; see e.g. \cite{erdmenger}. Given this set up, however,
we show that the holographic system
does undergo spontaneous symmetry breaking that adheres closely to the
BCS paradigm. We do so in a fluid limit
for the many-body fermion system where we explicitly construct the BCS
corrections to the fluid. The fluid limit has the advantage that  
we can compute the fully
backreacted gravitational solution and hence understand the
thermodynamic characteristics of the dual field theory.\footnote{See
  \cite{inprogress2} for a more microscopic study of pairing driven
  superconductivity in holography.}  The symmetry
breaking solution we find, therefore builds upon the Tolman-Oppenheimer-Volkov
self-gravitating 
Fermi fluid
solution underpinning neutron and electron stars. Indeed the BCS star
is readily recognizable as an AdS electron star cousin of an
astrophysical neutron star with a
superconducting core. We show that at zero temperature and with a positive
coupling, the corresponding BCS star solution is indeed the more
stable groundstate than the pure electron star solution. As a function
of the BCS coupling $\lambda$, the transition between the electron star and the
BCS star can be seen as an interaction driven (continuous) quantum phase
transition between the symmetry preserving state at $\lambda=0$ and
the symmetry-broken state at $\lambda\neq 0$. 

The symmetry breaking nature of the BCS star is confirmed by the
appearance of a pseudo-gap in the Fermi spectral function of the
boundary theory with the size of the gap is determined by the coupling
constant.
In addition the changes of the charge density at a fixed chemical potential %Luttinger's theorem is violated 
for a BCS star solution
implies the loss of charge in a superconducting state. Finally 
the conductivity is indeed suppressed at very low frequency, although
as is characteristic of holographic superconductors,
it does not exhibit a hard gap.

% One problem of this model is that we cannot see a hard gap in the optical conductivity at very small frequencies. This is related to the fact that the gap in the Fermi spectral function is also a pseudo-gap which only vanishes at zero frequency. 
A primary motivation of our work is to build a realistic holographic 
superconductor in that it explicitly encodes the fermionic degrees of freedom present in real exotic 
superconductors. On the gravity side of the duality, we 
show that considerations of what is natural there, gives a novel Stueckelberg-like coupling of the gap 
field. Interestingly, in the resulting BCS-Stueckelberg star, the susceptibility of the system towards 
superconductivity is enhanced, even though the suppression of the gap remains exponential.

There are various avenues to pursue to make the system even
more realistic. An obvious one is to consider lattice-effects and to 
encode the d-wave symmetry. In ordinary metals, the lattice phonons are responsible for the
effective four point interaction of the fermions. 
It is likely that the same will happen in a holographic
set-up with explicit fermions at finite density, as much of the fermionic physics follows the standard rules.
In that sense our BCS study here carries few surprises, but it serves
as another excellent benchmark of holographic duality. It also serves as
stepping stone. Using this BCS star as a base, an inquiry that tries
to connect it closer to the physics of strongly coupled physics that underly the
AdS/CFT duality could provide genuinely
new insights into the onset of superconductivity in quantum critical metals.
% Another future direction is to consider finite temperature states of this system and study the order of this phase transition.
%\comment{any comparison: fluid vs Hard wall BCS??}

%%%%%%%%%%%%%%%%%%%%%%%%%%%%%%%%%%
\appendix\section{Fluid parameters in region II}
\label{fluidapp}
%%%%%%%%%%%%%%%%%%%%%%%%%%%%%%%%%%

To obtain the result for the fluid parameters in region II quoted in
Eqs.~\eqref{IIa}-\eqref{IIc}, one subtracts the free fermion
contribution from region II, Eq. \eqref{FLII} from the formal
expressions
Eqs. \eqref{rho}, \eqref{eq:3}, and \eqref{p}. Using the $\omega_D
\ll \mu$ expansion
for the density of states in these differences, one obtains the
following expressions, where the
the integrations can be performed explicitly.
\begin{align}
  \label{eq:10}
n_{\text{II}}-n_{\text{II}}^{\text{FL}}  &\simeq \omega_D^2\nu_1 -\int_{-\ome_D}^{\ome_D} d\xi \frac{\xi^2\nu_1}{\sqrt{\xi^2+\Delta^2}}\non
&= -\nu_1
\frac{\Delta^2}{2} +\nu_1 \Delta^2 \ln\frac{2\ome_D}{\Delta}~,\non
\rho_{\text{II}} -\rho_{\text{II}}^{\text{FL}} 
&\simeq (\ome_D^2 -2\mu_l\ome_D)\nu_0 -(2\frac{\ome_D^3}{3}-\mu_l\ome_D^2)\nu_1-\nu_0\int_{-\ome_D}^{\ome_D} d\xi
  {\sqrt{\xi^2+\Delta^2}}+\mu_l n_{\text{II}}+ \rho_{\Delta}\non
&=-(\nu_0+\mu_l\nu_1)
  \frac{\Delta^2}{2} -(\nu_0-\mu_l\nu_1)\Delta^2 \ln\frac{2\ome_D}{\Delta}+\rho_{\Delta}~,\non
p_{\text{II}}-p_{\text{II}}^{\text{FL}} &\simeq\omega_D^2\nu_0 
-\frac{1}{3}\left[\nu_1\frac{\mu_l^2-m_f^2}{\mu}
  +\nu_0\left(\frac{-\mu_l^2+m_f^2}{\mu_l^2} +2\right)\right] \int_{-\ome_D}^{\ome_D} d\xi\frac{\xi^2}
  {\sqrt{\xi^2+\Delta^2}} +p_{\Delta}\non
&= 
-\nu_0\frac{\Delta^2}{2} +\nu_0\Delta^2\ln\frac{2\ome_D}{\Delta} +p_{\Delta}~.
\end{align}
Expanding the
integrated result in $\Delta \ll \ome_D$, while keeping the term
  $\rho_{\Delta}=-p_{\Delta}=\Delta^2/2\lambda$, one finds the expressions Eqs.~\eqref{IIa}-\eqref{IIc}.

%%%%%%%%%%%%%%%%%%%%%%%%%%%%%%%%%%%%%%%%%
\acknowledgments
%%%%%%%%%%%%%%%%%%%%%%%%%%%%%%%%%%%%%%%%%
We thank A. Bagrov, R.G. Cai, S. Gubser, G. Horowitz, K. Landsteiner, E. Lopez, B. Meszena and  S. Sachdev for discussions. 
This work was supported in part
by a VICI (KS) and a Spinoza grant (JZ) of the
Netherlands Organization for Scientific Research (NWO), by the
Netherlands Organization for Scientific Reseach/Ministry of Science
and Education (NWO/OCW) and by
% funding from
the Foundation for Research into Fundamental Matter (FOM) and by the support of the Spanish MINECO's ``Centro de Excelencia Severo Ochoa" Programme (YL and YWS) under grant SEV-2012-0249.

\end{document}